\documentclass[english,reprint, citeautoscript, prb,superscriptaddress]{revtex4-2}
\usepackage[T1]{fontenc}
\usepackage[utf8]{inputenc}
\usepackage{color}
\usepackage{verbatim}
\usepackage{amsmath}
\usepackage{amssymb}
\usepackage{graphicx}
\usepackage{wasysym}
\usepackage{babel}

\definecolor{niceblue}{RGB}{0,50,150}
\definecolor{nicered}{RGB}{210,50,0}
\usepackage[colorlinks,breaklinks,bookmarks=true,citecolor=blue,linkcolor=blue,urlcolor=black]{hyperref}

\newcommand{\im}{\mathrm{i}}
\newcommand{\mathd}{\mathrm{d}}

\begin{document}
\title{Langevin dynamics of generalized spins as SU($N$) coherent states}
\author{David Dahlbom}
\affiliation{Department of Physics and Astronomy, The University of Tennessee,
Knoxville, Tennessee 37996, USA}
\author{Cole Miles}
\affiliation{Department of Physics, Cornell University, Ithaca, New York 14850, USA}
\author{Hao Zhang}
\affiliation{Department of Physics and Astronomy, The University of Tennessee,
Knoxville, Tennessee 37996, USA}
\author{Cristian D. Batista}
\affiliation{Department of Physics and Astronomy, The University of Tennessee,
Knoxville, Tennessee 37996, USA}
\author{Kipton Barros}
\email{kbarros@lanl.gov}
\affiliation{Theoretical Division and CNLS, Los Alamos National Laboratory, Los
Alamos, New Mexico 87545, USA}

\begin{abstract}
Classical models of spin systems traditionally retain only the dipole
moments, but a quantum spin state will frequently have additional
structure. Spins of magnitude $S$ have $N=2S+1$ levels. Alternatively, the spin 
state is fully characterized by a set of $N^{2}-1$ local physical
observables, which we interpret as generalized spin components. For example,
a spin with $S=1$ has three dipole components and five quadrupole
components. These components evolve under a generalization of the classical Landau-Lifshitz dynamics, which can be extended with noise and damping terms. In this paper, we reformulate the dynamical equations of motion as a Langevin dynamics
of SU($N$) coherent states in the Schrödinger picture. This viewpoint is especially useful as
the basis for an efficient numerical method to sample spin configurations in thermal equilibrium and to simulate the relaxation and driven motion of topological solitons. To illustrate the approach, we simulate
a non-equilibrium relaxation process that creates CP$^{2}$ skyrmions,
which are topological defects with both dipole and quadrupole
character.
\end{abstract}
\maketitle

\section{Introduction}

The Landau-Lifshitz dynamics,

\begin{equation}
\frac{\mathd\mathbf{s}_{j}}{\mathd t}=-\mathbf{s}_{j}\times\frac{\partial H}{\partial\mathbf{s}_{j}},\label{eq:LLD}
\end{equation}
describes the evolution of classical spin dipoles $\mathbf{s}_{j}$, with site index $j$,
coupled through a conserved Hamiltonian $H(\mathbf{s}_{1},\dots,\mathbf{s}_{L})$.
The spin dipoles arise as expectations of the quantum spin operators.
States with spin $S\in\{\frac{1}{2},1,\frac{3}{2},\dots\}$ have $N=2S+1$
levels. When $S>1/2$, the local quantum state will contain multipolar
moments beyond the spin dipole, and our interest is modeling the dynamics
of these additional variables. A generalization of the Landau-Lifshitz
dynamics is especially important to model magnets with strong single-ion
anisotropy induced by the combination of spin-orbit coupling and crystal
field effects, such as 4\emph{d}--5\emph{d} and 4\emph{f}--5\emph{f}
electron materials as well as several $3d$ magnets~\cite{Zapf06,Do2020,Bai21}.
Eigenstates of single-ion anisotropy terms may have no net dipole moment, 
implying that higher order multipoles are needed for their characterization. Correspondingly, it is necessary to generalize the classical equations of motion  to account for the dynamics of these multipoles and for their effect on the time of evolution of the dipolar components, which can be accessed via inelastic neutron scattering experiments~\cite{Zhang21,Remund2022,Dahlbom22}.  
Such a generalization can be used to model lattices of locally entangled
units, such as of dimers~\cite{Jaime04},
trimers~\cite{Qiu05} and tetrahedra~\cite{Okamoto13}. Since each unit is an $N$-level system, one can take a classical limit based on SU($N$) coherent states, which captures the local entanglement of the unit, as well as the different excitation modes. For instance, the ground state of weakly-coupled antiferromagnetic spin 1/2 dimers ($N=4$) can be approximated by a direct product of singlet states, while the elementary excitations are triplon modes. Both aspects of the problem are captured by a classical dynamics of SU(4) coherent states. 

Consider a local spin with $N$ levels. It may
be characterized by its full set of expected multipole moments, which
consist of $N^{2}-1$ scalar components, $n_{j}^{\alpha}$. For example,
a spin with $S=1$ has $N=3$ levels, and gives rise to nontrivial
dipole and quadrupole moments, yielding a total of $3+5=3^{2}-1$
components in the vector $\mathbf{n}_{j}$. Using the classical approximation
that spins at different sites interact only through their expectation
values, one may derive a generalized spin dynamics~\cite{Zhang21,Dahlbom22},

\begin{equation}
\frac{\mathd\mathbf{n}_{j}}{\mathd t}=-\mathbf{n}_{j}\star\frac{\partial H}{\partial\mathbf{n}_{j}}.\label{eq:GSD}
\end{equation}
Here, we are defining the symbol $\star$ to denote the product
\begin{equation}
\left(\mathbf{a}\star\mathbf{b}\right)^{\alpha}=f_{\alpha\beta\gamma}a^{\beta}b^{\gamma},
\end{equation}
where $f_{\alpha\beta\gamma}$ are totally antisymmetric structure
constants of the $\mathfrak{su}(N)$ Lie algebra, and
summation over repeated Greek indices $\beta,\gamma$ is implied. In the special case of SU(2), $\star$ reduces to the usual vector cross product.

The full derivation of Eq.~(\ref{eq:GSD}) will be reviewed in Sec.~\ref{sec:review},
but we can briefly provide some intuition. Consider, first, a site $j$
with $N=2$ levels. Any Hamiltonian acting in the local two-dimensional
Hilbert space can be decomposed as a linear combination of the three
spin operators $\hat{S}_{j}^{\{x,y,z\}}$, plus a constant shift.
Each spin operator evolves according to its commutator with the Hamiltonian.
As generators for SU(2), the spin operators are defined to satisfy
the commutation relation $[\hat{S}_{j}^{\alpha},\hat{S}_{j}^{\beta}]=\im\,\epsilon_{\alpha\beta\gamma}\hat{S}_{j}^{\gamma}$,
with $\epsilon_{\alpha\beta\gamma}$ the Levi-Civita symbol. This commutator is
the source of the vector cross product in Eq.~(\ref{eq:LLD}), where
$s_{j}^{\alpha}=\langle\hat{S}_{j}^{\alpha}\rangle$. The generalization
to an $N$-level system is as follows. In place of $\hat{S}_{j}^{\alpha}$,
there are now $N^{2}-1$ operators $\hat{T}_{j}^{\alpha}$ that span
the full space of local physical observables. These are generators of
SU($N$), closed under commutation $[\hat{T}_{j}^{\alpha},\hat{T}_{j}^{\beta}]=\im\,f_{\alpha\beta\gamma}\hat{T}_{j}^{\gamma}$.
Repeating the same procedure as before, one arrives at Eq.~(\ref{eq:GSD}),
involving the $N^{2}-1$ expectation values $n_{j}^{\alpha}=\langle\hat{T}_{j}^{\alpha}\rangle$. Classical energy $H$ arises as the expectation of the quantum Hamiltonian.

In a real material at finite temperature, spins will be interacting
with additional degrees of freedom, such as lattice phonons, that
act as a thermal bath. To model such interactions implicitly, one
may include effective damping and noise terms into the dynamics. For
dipoles alone, this thermal coupling yields the stochastic Landau-Lifshitz
dynamics~\cite{Kubo70,Denisov94,Antropov96}, which has been the
subject of extensive study~\cite{Antropov97,Skubic08,Mentink10,Ma11,Ableidinger17}.

This paper investigates a generalized form of the stochastic spin dynamics,
\begin{equation}
\frac{\mathd\mathbf{n}_{j}}{\mathd t}=-\mathbf{n}_{j}\star\left(\boldsymbol{\xi}_{j}+\frac{\partial H}{\partial\mathbf{n}_{j}}-\lambda\mathbf{n}_{j}\star\frac{\partial H}{\partial\mathbf{n}_{j}}\right).\label{eq:S_GSD}
\end{equation}
The parameter $\lambda$ controls the damping magnitude. The $N^2-1$
components of $\boldsymbol{\eta}_{j}(t)$ are each Gaussian white noise,
defined by the moments
\begin{align}
\left\langle \xi_{j}^{\alpha}(t)\right\rangle  & =0\label{eq:xi1}\\
\left\langle \xi_{j}^{\alpha}(t)\xi_{k}^{\beta}(t')\right\rangle  & =2D\delta_{jk}\delta_{\alpha\beta}\delta(t-t'),\label{eq:xi2}
\end{align}
where
\begin{equation}
D=\lambda k_{B}T.\label{eq:D_lambda}
\end{equation}
One recovers the usual stochastic Landau-Lifshtz dynamics when restricting
to dipole moments, $\mathbf{n}_{j}\rightarrow\mathbf{s}_{j}$.

Assuming ergodicity, a fluctuation-dissipation theorem, derived in
Appendix~\ref{sec:fluctuation_dissipation}, ensures that the stochastic
dynamics samples from the Boltzmann equilibrium distribution $\propto\exp(-\beta H)$
for the classical Hamiltonian $H$, at inverse temperature $\beta=1/k_{B}T$.
Note that the noise term $\mathbf{n}_{j}\star\boldsymbol{\eta}_{j}$
is multiplicative, i.e., dependent on the dynamical spin variables $\mathbf{n}_{j}$;
this term must be interpreted in the sense of Stratonovich~\cite{Hasegawa80,vanKampen07}.

Equation~(\ref{eq:GSD}), for arbitrary Lie groups, is known in the
math literature as a Lie-Poisson system~\cite{Marsden13,Hairer06}.
The stochastic extension, Eq.~(\ref{eq:S_GSD}), has also recently
been studied~\cite{Arnaudon18}. Our quantum mechanical context,
however, brings additional structure. The fact that spin components
must be expectation values $n_{j}^{\alpha}=\langle Z_{j}|\hat{T}_{j}^{\alpha}|Z_{j}\rangle$
with respect to some underlying coherent state $|Z_{j}\rangle$ is
highly constraining. Given a basis, the expectation values
are $n_{j}^{\alpha}=\mathbf{Z}_{j}^{\dagger}T^{\alpha}\mathbf{Z}_{j}$,
where $\mathbf{Z}_{j}$ is an $N$-component complex vector representing
$|Z_{j}\rangle$, and matrices $T^{\alpha}$ are generators of SU($N$)
in the fundamental representation. Rather than working with the $N^{2}-1$
real components in $\mathbf{n}_{j}$, it is more economical to work
with the $N$ complex amplitudes in $\mathbf{Z}_{j}$.

A key contribution of our work is to reformulate Eq.~(\ref{eq:S_GSD})
as an evolution of the coherent states, $\mathbf{Z}_{j}(t)$. Our
final result, derived in Sec.~\ref{sec:stoch_schro}, is
\begin{equation}
\frac{\mathd\mathbf{Z}_{j}}{\mathd t}=-\im\,P_{j}\left[\boldsymbol{\zeta}_{j}+(1-\im\,\tilde{\lambda})\,\mathfrak{H}_{j}\mathbf{Z}_{j}\right],\label{eq:stoch_schro}
\end{equation}
where $\im$ is the unit imaginary number, and
$\tilde{\lambda}$ is a certain rescaling of $\lambda$. The operator
$P_{j}=I-\mathbf{Z}_{j}\mathbf{Z}_{j}^{\dagger}$ projects onto the
space orthogonal $\mathbf{Z}_{j}$, and therefore ensures unitary
evolution, i.e., conservation of the norm $|\mathbf{Z}_{j}|$. The $N$ components of the
vector $\boldsymbol{\zeta}_{j}$ are each complex Gaussian white noise, and scale like $\tilde{\lambda}^{1/2}$. Finally, the $N\times N$ matrices
\begin{equation}
\mathfrak{H}_{j}=\frac{\partial H}{\partial n_{j}^{\alpha}}T^{\alpha},\label{eq:H_frak}
\end{equation}
may be interpreted as local mean-field Hamiltonians that generate energy conserving time evolution of the coherent states $\mathbf{Z}_j$~\cite{Dahlbom22}.

Although the form of Eq.~\eqref{eq:stoch_schro} hints at a quantum mechanical origin, we emphasize that this dynamics is in fact derived by taking a classical limit. For example, the coherent states $\mathbf{Z}_j$ are not subject to any mean-field self-consistency constraint. To derive Eq.~\eqref{eq:stoch_schro}, we are exploiting the fact that the true Schrödinger dynamics of coherent states is independent of the particular degenerate irreducible representation of SU($N$). Correspondingly, the time evolution generated by $\mathfrak{H}_j$ coincides with  the classical dynamics, Eq.~\eqref{eq:GSD}. Note that the classical limit corresponds to replacing the
fundamental representation of SU($N$) with the degenerate irrep labeled by $\lambda_1 \to \infty$ and $\lambda_a =0$ for $2 \leq a \leq N-1$, where $\lambda_a$ are the eigenvalues of the generators of the Cartan subalgebra when applied to the maximal weight eigenstate~\cite{Zhang21} ($\lambda_1=1$ for the fundamental representation). In particular, this correponds to the $S\to \infty$ limit for the SU(2) case, where $\lambda_1 = 2S$. Since the Schrödinger dynamics of coherent states is independent of $\lambda_1$, even when we are working in the classical limit ($\lambda_1 \to \infty$), it is numerically favorable to use the fundamental irrep of SU($N$) in which the generators are the $N\times N$ matrices $T^\alpha$.

To summarize: Equation~\eqref{eq:GSD} is a classical approximation to the quantum many-body dynamics that neglects entanglement between sites, and generalizes the Landau-Lifshitz equation. Equation~\eqref{eq:S_GSD} adds to this phenomenological damping and noise terms. Equation~\eqref{eq:stoch_schro} is a mathematically equivalent reformulation of this stochastic dynamics; the information contained in the expected spin vector $\mathbf{n}_j$ is  more concisely captured by the underlying quantum coherent state $\mathbf Z_j$.

The Schrödinger picture is especially powerful for numerical simulation. Although structure constants $f_{\alpha\beta\gamma}$ appear explicitly in Eq.~\eqref{eq:S_GSD}, they are implicit in the equivalent Schrödinger dynamics of Eq.~\eqref{eq:stoch_schro}. In practice, one only needs to pick an explicit matrix basis for the spin operators, and the matrix $\mathfrak{H}$ will be a polynomial of these.
Selecting $\tilde{\lambda} = 0$ disables the Langevin damping and noise terms, and Eq.~(\ref{eq:stoch_schro})
becomes a canonical Hamiltonian system; here, the Schrödinger picture facilitates the
development of numerical integration methods that exactly conserve
the symplectic structure of the dynamics~\cite{Dahlbom22}. Finite thermal
coupling ($\tilde{\lambda}>0$) is useful
for sampling generalized spins in thermal equilibrium for the classical
Hamiltonian $H$.

Section~\ref{sec:examples} will demonstrate the formalism for a number of model systems, and compare with results obtained by numerical integration of Eq.~\eqref{eq:stoch_schro}. For single-site systems with spin $S > 1/2$ and strong anisotropy, multipolar spin states are necessary to capture all available degrees of freedom, and the physically correct dynamics. The formalism of SU($N$) coherent states also makes possible the treatment of entangled units, which we demonstrate by example. Finally, we will show the power of the method by performing large-scale simulations for a lattice of $S=1$ spins, competing exchange interactions, and a strong easy-axis anisotropy. Starting from an initially high temperature, a rapid nonequilibrium quench to low temperatures gives rise to a long-lived, metastable liquid of CP$^2$ skyrmions.

\section{Review of the generalized spin dynamics\label{sec:review}}

\subsection{Derivation of generalized spin dynamics as a Lie-Poisson system}

Let us now review the approximations leading to the generalized spin
dynamics, Eq.~(\ref{eq:GSD}). Our starting point is an arbitrary
quantum Hamiltonian. For example, a typical spin system might include
single-ion anisotropies and exchange interactions,
\begin{equation}
\hat{\mathcal{H}}_{\mathrm{spin}}=\sum_{j}f_{j}(\hat{\mathbf{S}}_{j})+\sum_{j,k}J_{(j,\alpha)(k,\beta)}\hat{S}_{j}^{\alpha}\hat{S}_{k}^{\beta},\label{eq:H_spin}
\end{equation}
with $f_{j}(\cdot)$ an arbitrary polyomial, and summation over repeated
Greek indices $\alpha$ and $\beta$ implied. If the local spin state
has magnitude $S$, then each local Hilbert space has dimension $N=2S+1$.
The space of local physical observables (up to a trace) is spanned by a set of $(N^{2}-1)$ traceless Hermitian operators $\hat{T}_{j}^{\alpha}$, that are generators for SU($N$). These generators form a basis of the $\mathfrak{s u}(N)$  Lie algebra and satisfy the commutation relations (Lie bracket)
\begin{equation}
[\hat{T}_{j}^{\alpha},\hat{T}_{j}^{\beta}]=\mathrm{i}\,f_{\alpha\beta\gamma}\hat{T}_{j}^{\gamma}.\label{eq:bracket_op}
\end{equation}
The structure constants $f_{\alpha\beta\gamma}$ depend on the choice of generators.

An arbitrary Hamiltonian up to quadratic order in the generators  may be written,
\begin{equation}
\hat{\mathcal{H}}=\sum_{j}J_{(j,\alpha)}^{(1)}\hat{T}_{j}^{\alpha}+\sum_{j,k}J_{(j,\alpha),(k,\beta)}^{(2)}\hat{T}_{j}^{\alpha}\hat{T}_{k}^{\beta}.\label{eq:H_many}
\end{equation}
This class of Hamiltonians includes  spin systems with arbitrary single-ion anisotropy term, Zeeman coupling to an external magnetic field and bilinear exchange interactions. 
Due to the completeness of the generators $\hat{T}_{j}^{\alpha}$, any single-ion anisotropy term can be expressed as a linear combination of the $\hat{T}_{j}^{\alpha}$. 
Therefore, without loss of generality, the interaction term only couples distinct sites,
\begin{equation}
J_{(j,\alpha),(j,\beta)}^{(2)}=0.\label{eq:J2_sym}
\end{equation}

Let $|\Psi(t)\rangle$ denote an arbitrary, time-evolving state. The expectation value of an observable $\hat{A}$ evolves as,
\begin{equation}
\mathrm{i}\,\frac{\mathd}{\mathd t}\langle \Psi| \hat{A} | \Psi \rangle=\langle \Psi |\, [\hat{A},\hat{\mathcal{H}}]|\,\Psi\rangle.\label{eq:A_dyn}
\end{equation}

In the classical limit  $\lambda_1 \to \infty$~\cite{Zhang21}, 
the wave function $|\Psi\rangle$ becomes a tensor product state
\begin{equation}
|Z\rangle=\bigotimes_{j}|Z_{j}\rangle.
\end{equation}
at all times $t$. Each local coherent state $|Z_{j}\rangle$ can be characterized by $N$
complex amplitudes in some basis, or alternatively, by the complete set
of expectation values,
\begin{equation}
n_{j}^{\alpha}=\langle Z_{j}|\hat{T}_{j}^{\alpha}|Z_{j}\rangle,\label{eq:n_def_op}
\end{equation}
for $\alpha=1,\dots,N^{2}-1$. Conversely, any such vector $\mathbf{n}_{j}$
corresponds to a unique coherent state $|Z_{j}\rangle$.

Classical states lack entanglement between distinct sites $j\neq k$, and expectation values factorize,
\begin{equation}
\langle\hat{T}_{j}^{\alpha}\hat{T}_{k}^{\beta}\rangle = \langle\hat{T}_{j}^{\alpha}\rangle\langle\hat{T}_{k}^{\beta}\rangle=n_{j}^{\alpha}n_{k}^{\beta}.
\end{equation}
This result, in combination with Eq.~(\ref{eq:J2_sym}), yields the expected energy for the classical state,
\begin{equation}
H=\langle \hat{\mathcal{H}}\rangle=\sum_{j}J_{(j,\alpha)}^{(1)}n_{j}^{\alpha}+\sum_{j,k}J_{(j,\alpha),(k,\beta)}^{(2)}n_{j}^{\alpha}n_{k}^{\beta}.\label{eq:H_classical}
\end{equation}
which has the same polynomial form as $\hat{\mathcal{H}}$. The function
$H(\mathbf{n}_{1},\dots,\mathbf{n}_{L})$ will also serve as the classical
Hamiltonian.

Equation~\eqref{eq:A_dyn}, along with the Hamiltonian in~\eqref{eq:H_many}, defines the time dynamics of arbitrary expectation values. Selecting $\hat{A}=\hat{T}_{j}^{\alpha}$ defines the time dynamics of $n_j^\alpha$. Under the assumed classical limit, $|\Psi\rangle \rightarrow |Z\rangle$, a short calculation yields (see Appendix B of Ref.~\onlinecite{Dahlbom22}),
\begin{equation}
\mathrm{i}\,\frac{\mathd n_{j}^{\alpha}}{\mathd t}=\langle[\hat{T}_{j}^{\alpha},\hat{T}_{j}^{\beta}]\rangle\frac{\partial H}{\partial n_{j}^{\beta}}.
\end{equation}
Inserting the Lie bracket of Eq.~(\ref{eq:bracket_op}) then produces the
generalized spin dynamics,
\begin{equation}
\frac{\mathd n_{j}^{\alpha}}{\mathd t}=f_{\alpha\beta\gamma}\frac{\partial H}{\partial n_{j}^{\beta}}n_{j}^{\gamma}.\label{eq:GSD-2}
\end{equation}
This takes the standard form for a Lie-Poisson system~\cite{Marsden13},
and holds for an arbitrary choice of SU($N$) generators $\hat{T}_{j}^{\alpha}$.
The final result would be unchanged had we included three-body or
higher-order couplings in the Hamiltonian, $\hat{\mathcal{H}}$. Imposing
an orthonormality condition on the generators $\hat{T}_{j}^{\alpha}$
ensures that $f_{\alpha\beta\gamma}$ is antisymmetric in all indices,
and makes contact with the generalized spin dynamics, Eq.~(\ref{eq:GSD}).

Although we presented this discussion in the context of a quantum
spin Hamiltonian, Eq.~(\ref{eq:H_spin}), the final result is fully
general. The many-body Hamiltonian of Eq.~(\ref{eq:H_many}) could
be used to model \emph{any} quantum system that couples local $N$-level
degrees of freedom, which need not have a spin character.

\subsection{Mapping to a mean-field Schrödinger equation\label{subsec:schrodinger_derivation}}

Here we take a direct path to derive the main result of Ref.~\onlinecite{Dahlbom22},
which reformulates Eq.~(\ref{eq:GSD-2}) as a Schrödinger dynamics
of coherent state vectors.

Let us first establish some notation. Assume some fixed basis $\{|e_{1}\rangle,\dots,|e_{N}\rangle\}$
such that each local coherent state $|Z_{j}\rangle$ becomes a vector
$\mathbf{Z}_{j}$ containing $N$ complex amplitudes, 
$
Z_{j,a}=\langle e_{a}|Z_{j}\rangle.
$
Similarly, each generator $\hat{T}_{j}^{\alpha}$ becomes an $N\times N$
Hermitian matrix $T^{\alpha}$, independent of the site index $j$. Equation~(\ref{eq:n_def_op})
for the local spin components becomes 
\begin{equation}
n^{\alpha}_{j}=\mathbf{Z}_{j}^{\dagger}T^{\alpha}\mathbf{Z}_{j}.\label{eq:n_def}
\end{equation}
The commutation relation of Eq.~(\ref{eq:bracket_op}) still holds,
\begin{equation}
[T^{\alpha},T^{\beta}]=\im\,f_{\alpha\beta\gamma}T^{\gamma}.\label{eq:bracket}
\end{equation}
We will require the SU($N$) generators to be orthonormal,
\begin{equation}
\mathrm{tr}\,T^{\alpha}T^{\beta}=\tau\delta_{\alpha,\beta}.\label{eq:Lie_ortho}
\end{equation}
The constant $\tau$ is determined by the convention for the overall
magnitude of the generators,
\begin{equation}
\tau=\mathrm{tr}\,T^{\alpha}T^{\alpha}\equiv\left\Vert T^{\alpha}\right\Vert ^{2},\label{eq:tau_def}
\end{equation}
independent of $\alpha$ (no sum implied here). Orthonormality ensures that the structure
constants are totally antisymmetric,
\begin{equation}
f_{\alpha\beta\gamma}=-f_{\beta\alpha\gamma}=-f_{\alpha\gamma\beta}.\label{eq:antisym}
\end{equation}
For simplicity, we employ the convention that coherent states are
normalized to unity,
\begin{equation}
\mathbf{Z}_{j}^{\dagger}\mathbf{Z}_{j}=1.
\end{equation}

With this notation established, now consider the outer product,
\begin{equation}
\rho_{j}=\mathbf{Z}_{j}\mathbf{Z}_{j}^{\dagger},\label{eq:rho_def}
\end{equation}
in analogy with the density matrix for the ``pure state'' $\mathbf{Z}_{j}$.
The cyclic property of the trace ensures
\begin{equation}
\mathrm{tr}\,\rho_{j}T^{\alpha}=n_{j}^{\alpha},\label{eq:rho_to_n}
\end{equation}
for arbitrary $\alpha$. As a Hermitian matrix, $\rho_{j}$ can always
be decomposed as a linear combination of generators $T^{\alpha}$,
plus a constant shift. By the orthonormality condition of Eq.~(\ref{eq:Lie_ortho}),
this decomposition must be
\begin{equation}
\rho_{j}=\frac{1}{\tau}n_j^{\alpha}T^{\alpha}+I,\label{eq:rho_expand}
\end{equation}
with $I$ the $N\times N$ identity, and summation over $\alpha$
implied. Note that $\mathfrak{n} = n^\alpha T^\alpha$ is known in high-energy physics as the color field. The dynamics of spin components, Eq.~(\ref{eq:GSD-2}), fixes the dynamics of $\rho_j$,
\begin{equation}
\frac{\mathd\rho_{j}}{\mathd t}=\frac{1}{\tau}\left(f_{\alpha\beta\gamma}\frac{\partial H}{\partial n_{j}^{\beta}}n_{j}^{\gamma}\right)T^{\alpha}.
\end{equation}
Total antisymmetry, Eq.~(\ref{eq:antisym}), implies $f_{\alpha\beta\gamma}=f_{\beta\gamma\alpha}$,
which allows to substitute Eq.~(\ref{eq:bracket}), yielding
\begin{equation}
\frac{\mathd\rho_{j}}{\mathd t}=-\frac{\im}{\tau}\left[T^{\beta},T^{\gamma}\right]\frac{\partial H}{\partial n_{j}^{\beta}}n_{j}^{\gamma}.
\end{equation}
Now undo the expansion of Eq.~(\ref{eq:rho_expand}) to find
\begin{equation}
\frac{\mathd\rho_{j}}{\mathd t}=-\im\left[\mathfrak{H}_{j},\rho_{j}\right],\label{eq:drho_dt}
\end{equation}
where $\mathfrak{H}_{j}=(\partial H/\partial n_{j}^{\alpha})T^{\alpha}$
was defined in Eq.~(\ref{eq:H_frak}). This dynamics may be interpreted
as the von Neumann evolution of a density matrix. Here, $\mathfrak{H}_{j}$
serves as a local quantum Hamiltonian for site $j$, and couples to
distinct sites $k$ via their expectation values $\mathbf{n}_{k}$.

One may verify by direct calculation that Eq.~(\ref{eq:drho_dt})
is satisfied if we take the coherent states to evolve as
\begin{equation}
\frac{\mathd\mathbf{Z}_{j}}{\mathd t}=-\im\,(\mathfrak{H}_{j}+cI)\mathbf{Z}_{j}.\label{eq:schrodinger}
\end{equation}
where $c(t)$ is an arbitrary, time-dependent shift of the Hamiltonian, representing a gauge freedom. Any trajectory $\mathbf{Z}_{j}(t)$ satisfying Eq.~(\ref{eq:schrodinger})
will also yield a trajectory $\mathbf{n}_{j}(t)$ satisfying the generalized
spin dynamics, Eq.~(\ref{eq:GSD-2}), which achieves our goal.

Let $\mathbf{Z}_{j}^{0}(t)$ denote an integrated trajectory using the familiar choice of gauge, $c(t)=0$. An alternative choice $c(t)$ would introduce a physically irrelevant complex phase, $\mathbf{Z}_{j}(t)=e^{-\im\,\theta(t)}\mathbf{Z}_{j}^{0}(t)$,
where $\theta(t)=\int_{0}^{t}c(t')\,\mathd t'$. This complex phase has no effect on $\rho_{j}=\mathbf{Z}_{j}\mathbf{Z}_{j}^{\dagger}$,
nor on expectation values $n_{j}^{\alpha}=\mathrm{tr}\,\rho_{j}T^{\alpha}$. Nonetheless, a careful choice of gauge can be helpful to simplify certain calculations.

Our results in Sec.~\ref{sec:stoch_schro} take their most elegant form using
$c(t)=-\mathbf{Z}_{j}^{\dagger}\mathfrak{H}_{j}\mathbf{Z}_{j}$, which corresponds to the parallel transport gauge~\cite{Lin2020}. With this choice, Eq.~(\ref{eq:schrodinger}) may be written
\begin{align}
\frac{\mathd\mathbf{Z}_{j}}{\mathrm{d}t} & =-\im\,P_{j}\mathfrak{H}_{j}\mathbf{Z}_{j},\label{eq:schrodinger_proj}
\end{align}
where the $N\times N$ matrix
\begin{equation}
P_{j}=I-\mathbf{Z}_{j}^{\dagger}\mathbf{Z}_{j},\label{eq:P_def}
\end{equation}
projects onto the vector subspace orthogonal to $\mathbf{Z}_{j}$.
The parallel transport gauge effectively minimizes the introduction
of complex phase throughout the trajectory $\mathbf{Z}(t)$.

The Schrödinger picture is numerically expedient for two reasons. First, it is a canonical Hamiltonian system; the dynamics satisfies Hamilton's equations of motion where the real and imaginary parts of $\mathbf{Z}_j$ act as canonical momenta and positions. This facilitates the design of symplectic integration schemes~\cite{Dahlbom22} and makes contact with related work in the math literature~\cite{McLachlan15,McLachlan14,Modin20}. Second, when $N \geq 3$, coherent states $\mathbf{Z}_{j}$ are the most concise representation of the actual information contained within the expectation values $\mathbf{n}_j$. Each coherent state $\mathbf{Z}_{j}$ is an $N$-component complex vector subject to a normalization constraint, and defined up to an overall complex phase. That is, $\mathbf{Z}_{j}$ lives in CP$^{N-1}$, a space with $2(N-1)$ real degrees of freedom. In contrast, the spin vector $\mathbf{n}_j$ requires $N^{2}-1$ components to capture the same underlying information.

\section{Stochastic spin dynamics in the Schrödinger picture \label{sec:stoch_schro}}

In this section we will reformulate the stochastic spin
dynamics as a dynamics of the local coherent states $\mathbf{Z}_{j}(t)$
in the Schrödinger picture.

Repeating the same procedure leading to Eq.~(\ref{eq:schrodinger_proj}),
the stochastic dynamics of Eq.~(\ref{eq:S_GSD}) can be mapped to
\begin{equation}
\frac{\mathd}{\mathd t}\mathbf{Z}_{j}=-\im\,P_{j}(a_{j}^{\alpha}T^{\alpha})\mathbf{Z}_{j}.
\end{equation}
where the coefficients
\begin{equation}
\mathbf{a}_{j}=\boldsymbol{\xi}_{j}+\frac{\partial H}{\partial\mathbf{n}_{j}}-\lambda\mathbf{n}_{j}\star\frac{\partial H}{\partial\mathbf{n}_{j}}
\end{equation}
now include noise, energy gradient, and damping parts. We decompose
the new effective Hamiltonian into three parts,

\begin{equation}
a_{j}^{\alpha}T^{\alpha}=\mathfrak{X}_{j}+\mathfrak{H}_{j}-\lambda\mathfrak{A}_{j}.
\end{equation}
The matrix $\mathfrak{H}_{j}$ was introduced in Eq.~(\ref{eq:H_frak}).
Also present is a Hermitian noise matrix,

\begin{equation}
\mathfrak{X}_{j}=\xi_{j}^{\alpha}T^{\alpha},\label{eq:frakX}
\end{equation}
and a matrix associated with damping,
\begin{equation}
\mathfrak{A}_{j}=f_{\alpha\beta\gamma}n_{j}^{\beta}\frac{\partial H}{\partial n_{j}^{\gamma}}T^{\alpha}.
\end{equation}

Using Eqs.~(\ref{eq:bracket}) and~(\ref{eq:antisym}), the latter
becomes
\begin{align*}
\mathfrak{A}_{j} & =-\im\,[T^{\beta},T^{\gamma}]n_{j}^{\beta}\frac{\partial H}{\partial n_{j}^{\gamma}}.
\end{align*}
Substituting Eqs.~(\ref{eq:rho_expand}) and~(\ref{eq:H_frak}),
\begin{equation}
\mathfrak{A}_{j}=-\im\,\tau[\rho_{j},\mathfrak{H}_{j}].
\end{equation}
Collecting results,

\begin{equation}
\frac{\mathd}{\mathd t}\mathbf{Z}_{j}=-\im\,P_{j}(\mathfrak{X}_{j}+\mathfrak{H}_{j}+\im\,\tilde{\lambda} [\rho_{j},\mathfrak{H}_{j}])\mathbf{Z}_{j},\label{eq:dZ_dt_1}
\end{equation}
where
\begin{equation}
\tilde{\lambda}=\tau\lambda.
\end{equation}
denotes a rescaling of the damping magnitude, with $\tau$ defined
in Eq.~(\ref{eq:tau_def}).

The noise and damping terms can be further simplified. Substituting
$\rho_{j}=\mathbf{Z}_{j}\mathbf{Z}_{j}^{\dagger}$ and using the normalization
condition $\mathbf{Z}_{j}^{\dagger}\mathbf{Z}_{j}=1$, we calculate
\begin{equation}
[\rho_{j},\mathfrak{H}_{j}]\mathbf{Z}_{j}=-P_{j}\mathfrak{H}_{j}\mathbf{Z}_{j},\label{eq:rho_frakh_commutator}
\end{equation}
where $P_{j}=I-\mathbf{Z}_{j}\mathbf{Z}_{j}^{\dagger}$ appears once
more. Using the idempotency property $P_{j}^{2}=P_{j}$,
\begin{align}
\frac{\mathd}{\mathd t}\mathbf{Z}_{j} & =-\im\,P_{j}(\mathfrak{X}_{j}+\mathfrak{H}_{j}-\im\,\tilde{\lambda}\mathfrak{H}_{j})\mathbf{Z}_{j}.\label{eq:s-gsd-intermediate}
\end{align}

For the noise term, Appendix~\ref{sec:noise} gives the result
\begin{equation}
P_{j}\mathfrak{X}_{j}\mathbf{Z}_{j}=P_{j}\boldsymbol{\zeta}_{j},\label{eq:eta_dot_z}
\end{equation}
where $\boldsymbol{\zeta}_{j}$ is a complex Gaussian white noise
vector. Its components have zero mean and second moment
\begin{equation}
\left\langle \zeta_{j,a}^{\ast}(t)\zeta_{k,b}(t)\right\rangle =2\tau D\delta_{j,k}\delta_{a,b}\delta(t-t').
\end{equation}
Recall that Eq.~(\ref{eq:D_lambda}) defines $D=\lambda k_\mathrm{B}T$.

This confirms the stochastic Schrödinger dynamics as stated in Eq.~(\ref{eq:stoch_schro}),
\[
\frac{\mathd}{\mathd t}\mathbf{Z}_{j}=-\im\,P_{j}\left[\boldsymbol{\zeta}_{j}+(1-\im\,\tilde{\lambda})\,\mathfrak{H}_{j}\mathbf{Z}_{j}\right],
\]
where
\begin{equation}
\left\langle \zeta_{j,a}^{\ast}(t)\zeta_{k,b}(t)\right\rangle =2\tilde{\lambda}k_\mathrm{B}T\delta_{j,k}\delta_{a,b}\delta(t-t').
\end{equation}
Observe that all factors of $\tau$ have effectively been absorbed
into a rescaling of the empirical damping magnitude, $\lambda\rightarrow\tilde{\lambda}$. 

The second-order Heun scheme, followed by a normalization step, is a convenient method for numerical integration of the stochastic dynamics. Details are provided in Appendix~\ref{sec:heun}.

\section{Applications \label{sec:examples}}

To illustrate the stochastic generalized spin dynamics, we will present a sequence of examples. To verify correctness of equilibrium statistics, we compare exact analytical results, Sec.~\ref{subsec:single-site}, with numerical integration, Secs.~\ref{sec:half_zeeman}--\ref{sec:dimer}, for several simple models. Finally, in Sec.~\ref{sec:skyrmion} we apply Langevin dynamics to study a large-scale nonequilibrium quench process, leading to the formation of CP$^2$ skyrmions.

The utility of the generalized spin dynamics is that it allows an
effective mean-field decoupling between distinct sites, or more generally, entangled units.  Such an entangled unit would still be modeled with a thermally fluctuating SU($N$) coherent state, but $N$ now represents the dimension of an expanded local Hilbert space involving the tensor product space of ``microscopic'' sites (e.g., sites in a dimer or trimer). In this way, the classical formalism of SU($N$) coherent states can also be used to model {\it local} quantum entanglement. This is particularly important because it enables a classical description of quantum systems where magnetic ordering is completely or partially suppressed due to strong anisotropy or singlet formation. Systems where the magnetic ordering is completely suppressed are known as quantum paramagnets. 

In the limit that deviations from the ground state are small, the classical dynamics is well approximated by uncoupled harmonic oscillators associated with each normal mode (small oscillations approximation). In quantum mechanical language, the quantization of each harmonic oscillator leads to an effective Hamiltonian quadratic in Schwinger boson operators. That is, our classical dynamics can be understood as an extension of linear spin wave theory, as generalized to SU($N$) coherent states~\cite{muniz2014}. 
Quantizing in this way, a classical-to-quantum correspondence prefactor of $\omega/kT$ appears in the correlation function. This prefactor may be applied as a correction to the correlation function, e.g.
as applied in Refs.~\onlinecite{ZhangChanglani2019, Remund2022}. We leave examination of this correction to a future study and focus here on the exact behavior of the classical models.

\subsection{Models with a single local Hilbert space\label{subsec:single-site}}

To demonstrate the formalism, we will begin with the simplified case of a
single site.
Recall that the most general quantum Hamiltonian can be written in
the form of Eq.~(\ref{eq:H_many}). By restricting to a single site
(local Hilbert space of dimension $N$), the Hamiltonian
\begin{equation}
\hat{\mathcal{H}}=J_{\alpha}\hat{T}^{\alpha},\label{eq:H_qm_single}
\end{equation}
becomes linear $N^{2}-1$ local operators $\hat{T}^{\alpha}$, interpreted
as orthonormal generators of SU($N$). Given a basis, the Schrödinger equation
\begin{equation}
        \frac{\mathd |Z\rangle}{\mathd t}=-\im\,\hat{\mathcal{H}} |Z\rangle,\label{eq:schrodinger-qm}
\end{equation}
becomes a dynamics of the complex vector $\mathbf{Z}$,
\begin{equation}
\frac{\mathd\mathbf{Z}}{\mathd t}=-\im\,\mathfrak{H}\mathbf{Z},\label{eq:schrodinger_single}
\end{equation}
where $\mathfrak{H}$ is the matrix representation of $\hat{\mathcal{H}}$. The linearity of the single-site Hamiltonian, Eq.~\eqref{eq:H_qm_single}, implies $\mathfrak{H}=(\partial H/\partial n_{j}^{\alpha}) T^{\alpha}$ where $H = J_{\alpha}n^{\alpha}$ is a special case of Eq.~\eqref{eq:H_classical}. It follows that Eq.~\eqref{eq:schrodinger_single} is a special case of Eq.~(\ref{eq:schrodinger}). The arguments in Sec.~\ref{subsec:schrodinger_derivation} therefore establish equivalence between the quantum and classical dynamics [Eqs.~\eqref{eq:schrodinger-qm} and~(\ref{eq:GSD}) respectively] for this single site model.

Another perspective on Eq.~\eqref{eq:schrodinger_single} is that it may be viewed as a classical dynamics of SU($N$) coherent states. Recall that this classical limit is obtained by taking the label $\lambda_1$ of the degenerate irrep of SU($N$) to infinity~\cite{Zhang21}.
In this limit, the coherent states become orthogonal to each other (note that the dimension of the vector space diverges in that limit) and quantum mechanical operators can be replaced by their expectation values.  Although a given coherent state has different representations, its time evolution is independent of the representation. This explains why the classical equation of motion, obtained for $\lambda_1 \to \infty$, can be mapped into the  Schrödinger equation~\eqref{eq:schrodinger-qm} in the original representation $\lambda_1=1$.

An approximation does arise, however, when calculating finite temperature
expectation values in the classical limit using SU($N$) coherent states. For reference, the quantum mechanically correct partition function is given by a trace over basis states,
\begin{equation}
\mathcal{Z}_{\mathrm{quantum}}=\mathrm{tr}\,e^{-\beta\mathcal{\hat{H}}}=\sum_{a=1}^{N}e^{-\beta\epsilon_{a}},\label{eq:Z_qm_single}
\end{equation}
where $\epsilon_{a}$ are the eigenvalues of $\hat{\mathcal{H}}$. Physical observables are defined similarly as traces.

In contrast, the stochastic Schrödinger
equation,~\eqref{eq:stoch_schro}, samples the continuous space of coherent states $\mathbf Z$ from a {\it classical} Boltzmann distribution
\begin{equation}
P(\mathbf{Z})\propto e^{-\beta H(\mathbf{Z})}.
\end{equation}
The expected energy $H=\mathbf{Z}^{\dagger}\mathfrak{H}\mathbf{Z}$ serves as the classical Hamiltonian. The corresponding partition function is,
\begin{equation}
\mathcal{Z}_{\textrm{SU}(N)} = \int_{\mathrm{CP}^{N-1}}e^{-\beta\mathbf{Z}^{\dagger}\mathfrak{H}\mathbf{Z}}\,\mathd\mathbf{Z},
\end{equation}
where the domain of integration is the complex projective space CP$^{N-1}$. Alternatively, up to an irrelevant scaling factor, we can integrate every component $Z_a$ of $\mathbf Z$ over the full complex plane, subject to the normalization constraint $|\mathbf Z|=1$,
\begin{equation}
\mathcal{Z}_{\textrm{SU}(N)}\propto\int_{\mathbb{C}^{N}}e^{-\beta\mathbf{Z}^{\dagger}\mathfrak{H}\mathbf{Z}}\delta(|\mathbf{Z}|^{2}-1)\,\mathd\mathbf{Z}.\label{eq:Z_cl_single}
\end{equation}
In this context, $\delta(|\mathbf{Z}|-1) \propto \delta(|\mathbf{Z}|^{2}-1)$.

This integral over classical coherent states can be evaluated exactly. Because we are working with a single site, the matrix $\mathfrak{H}$ is an exact representation of
the operator $\hat{\mathcal{H}}$, and has the same eigenvalues. The integral over $\mathbf{Z}$
is invariant to a unitary change of basis. Without loss of generality,
we may work in the eigenbasis of $\mathfrak{H}$, such that
\begin{equation}
\mathbf{Z}^{\dagger}\mathfrak{H}\mathbf{Z}=\sum_{a=1}^{N}\epsilon_{a}|Z_{a}|^{2}.
\end{equation}
By writing the components of $\mathbf{Z}$ in polar coordinates, $Z_{a}=x_{a}e^{\im\,\phi_{a}}$,
integrals over $\mathbb{C}$ may be replaced by integrals over $\mathbb{R}^{2}$,
where $\mathd Z_{a}\rightarrow x_{a}\mathd\phi_{a}\mathd x_{a}$.
In the eigenbasis of $\mathfrak{H}$, the integrals over phase $\phi_{a}$
are irrelevant, up to a scaling factor. It is convenient to change integration variables $y_{a}=x_{a}^{2}$ such that $x_a \mathd x_a \rightarrow \frac{1}{2} \mathd y_a$ and
\begin{equation}
\mathcal{Z}_{\textrm{SU}(N)}\propto\int e^{-\beta\sum_{a=1}^{N}\epsilon_{a}y_{a}}\delta(|\mathbf{Z}|^{2}-1)\,\mathd y_{1}\dots\mathd y_{N},
\end{equation}
where $|\mathbf{Z}|^{2}=y_{1}+\dots+y_{N}$. Integration over $y_{N}$
yields the substitution rule $y_{N}\rightarrow1-\sum_{a=1}^{N-1}y_{a}$.
The remaining $N-1$ integrals are,
\begin{equation}
\mathcal{Z}_{\textrm{SU}(N)}\propto e^{-\beta \epsilon_{N}}\int\! e^{-\beta\sum_{a=1}^{N-1}(\epsilon_{a}-\epsilon_{N})y_{a}} \,\mathd y_{1} \dots \mathd y_{N-1},\label{eq:Z_gs}
\end{equation}
where the integration domain is defined by the constraints $y_a \geq 0$ and $y_1 + \dots + y_{N-1} \leq 1$. One can select, e.g., $y_1 \in [0, 1]$, $y_2 \in [0, 1 - y_1]$, $y_3 \in [0, 1 - y_1 - y_2]$, and so forth.

Formal integration yields a result that is manifestly symmetric under permutation of eigenvalues,
\begin{equation}
\mathcal{Z}_{\mathrm{SU}(N)} \propto \sum_{a=1}^{N}  \frac{e^{-\beta\epsilon_{a}}}{\prod_{b \neq a} \beta (\epsilon_{b}-\epsilon_{a})} .\label{eq:Z_gs_integrated}
\end{equation}
The product $\prod_{b \neq a}$ runs over indices $b = 1,\dots,N$, excluding $b = a$. If degenerate eigenvalues are present, then some denominators of Eq.~\eqref{eq:Z_gs_integrated} will vanish. For example, if $\epsilon_1 = \epsilon_2$, then the first two terms in the sum are individually divergent. Note, however, that this apparent singularity is removable through algebraic manipulations, and appropriate cancellations. Alternatively, for a given model, it may be more convenient to use the original integral form, Eq.~\eqref{eq:Z_gs}.

The distinction between the quantum partition function, Eq.~\eqref{eq:Z_qm_single}, and its classical approximation, Eq.~(\ref{eq:Z_gs}), will be illustrated
with examples.

\subsection{Single spin-1/2 site with Zeeman coupling\label{sec:half_zeeman}}

The simplest example of the formalism is a single site with spin $S=1/2$
and Zeeman coupling. The Hamiltonian
\begin{equation}
\hat{\mathcal{H}}=-B\hat{S}^{z}
\end{equation}
has $N=2$ eigenvalues $\epsilon_{\{1,2\}}=\pm B/2$. The quantum
partition function is therefore

\begin{equation}
\mathcal{Z}_{\mathrm{quantum}}=e^{+\beta B/2}+e^{-\beta B/2}.\label{eq:Z_qm_zeeman}
\end{equation}
The classical partition function is given by Eq.~(\ref{eq:Z_cl_single})
or equivalently Eq.~(\ref{eq:Z_gs}),
\begin{equation}
\mathcal{Z}_{\textrm{SU}(2)}\propto e^{+\beta B/2}\int_{0}^{1}\mathd y_{1}\,e^{-\beta By_{1}}\propto\frac{1}{\beta B}\sinh(\beta B/2).\label{eq:Z_cl_zeeman}
\end{equation}

As a pedagogical exercise, we will now follow the more traditional
path of calculating $\mathcal{Z}_{\textrm{SU}(2)}$ as an integral
over expected spin components. Because the system has $N=2$ levels,
the three spin operators $\hat{S}^{\alpha}$ serve as a complete set
of orthonormal generators $\hat{T}^{\alpha}$, such that the generalized
spin $\mathbf{n}$ includes only a dipole part, $s^{\alpha}=\langle Z|\hat{S}^{\alpha}|Z\rangle$.
The expected energy,
\begin{equation}
H=\langle Z|\hat{\mathcal{H}}|Z\rangle=-Bs^{z},
\end{equation}
acts as the classical Hamiltonian. By the Bloch sphere construction,
the coherent states $|Z\rangle$ map isometrically to spin dipoles
$\mathbf{s}$ of magnitude $1/2$. The associated partition function
is,
\begin{equation}
\mathcal{Z}_{\textrm{SU}(2)}=\int_{\mathbb{R}^{3}}e^{+\beta Bs^{z}}\delta\left(|\mathbf{s}|-\frac{1}{2}\right)\mathd\mathbf{s}.\label{eq:Z_cl_zeeman_2}
\end{equation}
In spherical coordinates, $\mathd\mathbf{s}=s^{2}\mathd s\,\mathd\cos\theta\,\mathd\phi,$ where
$|\mathbf{s}|=s$, and $s^{z}=s\cos\theta$. Under a change
of integration variable $\cos\theta\rightarrow c$, the integral
\[
\mathcal{Z}_{\textrm{SU}(2)}\propto\int_{-1}^{+1}e^{+\beta Bc/2}\mathd c\propto\frac{1}{\beta}\sinh(\beta B/2)
\]
reproduces Eq.~(\ref{eq:Z_cl_zeeman}), as expected.

The associated mean energies are given by $E=-\partial\ln\mathcal{Z}/\partial\beta$,
with the results
\begin{align}
E_{\mathrm{quantum}} & =-\frac{B}{2}\tanh(\beta B/2)\\
E_{\textrm{SU}(2)} & =\beta^{-1}-\frac{B}{2}\coth(\beta B/2).\label{eq:E_cl_zeeman}
\end{align}
At small temperatures $k_\mathrm{B}T=\beta^{-1}$, the classical energy $E_{\textrm{SU}(2)}\sim\beta^{-1}-\frac{B}{2}$
grows linearly with temperature. This is an unphysical but well-known
limitation of the classical approximation to the thermal distribution
of quantum spin states. The correct energy, $E_{\mathrm{quantum}}$,
is approximately constant for temperatures $k_\mathrm{B}T$ much smaller
than the gap (energy scale $B$).

\subsection{Single spin-1 site with anisotropy}

As a next example, consider the Hamiltonian with spin $S=1$ and an
easy-axis anisotropy,
\begin{equation}
\hat{\mathcal{H}}=D\left(\hat{S}^{z}\right)^{2}.\label{eq:calH_aniso}
\end{equation}
The $N=3$ eigenvalues $\{-1,0,1\}$ of $\hat{S}^{z}$ give rise to
the eigenvalues $\{D,0,D\}$ of $\hat{\mathcal{H}}$. The quantum
partition function is
\begin{equation}
\mathcal{Z}_{\mathrm{quantum}}=1+2e^{-\beta D}.
\end{equation}

Generalized spin $\mathbf{n}$ now has $N^{2}-1=8$ total components.
Three are the usual expected dipole components,
\begin{equation}
s^{\alpha}=\langle Z|\hat{S}|Z\rangle.
\end{equation}
In addition, the quadrupole moments are defined via,
\begin{equation}
Q^{\alpha\beta}=\langle Z|\Big(\hat{S}^{\alpha}\hat{S}^{\beta}+ \hat{S}^{\beta}\hat{S}^{\alpha} - \frac{4}{3} \delta_{\alpha \beta}\Big)|Z\rangle,
\end{equation}
Note that $Q^{\alpha\beta}$ is symmetric and 
traceless, which leaves five quadrupole degrees of freedom.

In principle, the classical SU(3) partition function may be calculated
by directly integrating over the allowed dipole and quadrupole moments
contained within $\mathbf{n}$. In practice, it is much easier to
integrate over coherent states $|Z\rangle$, as in Eq.~(\ref{eq:Z_cl_single}).
Applying the result of Eq.~(\ref{eq:Z_gs}) with eigenvalues $\epsilon_{1}=\epsilon_{2}=D$
and $\epsilon_{3}=0$, the partition function is
\begin{align}
\mathcal{Z}_{\textrm{SU}(3)} & \propto\int_{0}^{1}\mathd y_{1}\int_{0}^{1-y_{1}}\mathd y_{2}\,e^{-\beta(Dy_{1}+Dy_{2})}.
\end{align}

An alternative, but more traditional, classical limit would replace
each coherent state by its expected spin dipole of magnitude 1. Here,
the classical energy,
\begin{equation}
H_{\mathrm{dipole}}=D\left(s^{z}\right)^{2},\label{eq:H_dipole}
\end{equation}
is quadratic in the expected spin. The associated partition function
\begin{equation}
\mathcal{Z}_{\mathrm{dipole}}=\int_{\mathbb{R}^{3}}e^{-\beta D(s^{z})^{2}}\delta\left(|\mathbf{s}|-1\right)\mathd\mathbf{s},
\end{equation}
can be integrated in spherical coordinates using $c=\cos\theta$,
\begin{align}
\mathcal{Z}_{\mathrm{dipole}} & \propto \int_{-1}^{+1}e^{-\beta Dc^{2}}\mathd c=\frac{\sqrt{\pi}\,\mathrm{erf}(\sqrt{D\beta})}{\sqrt{D\beta}}.
\end{align}

\begin{figure}

\includegraphics[width=0.9\columnwidth]{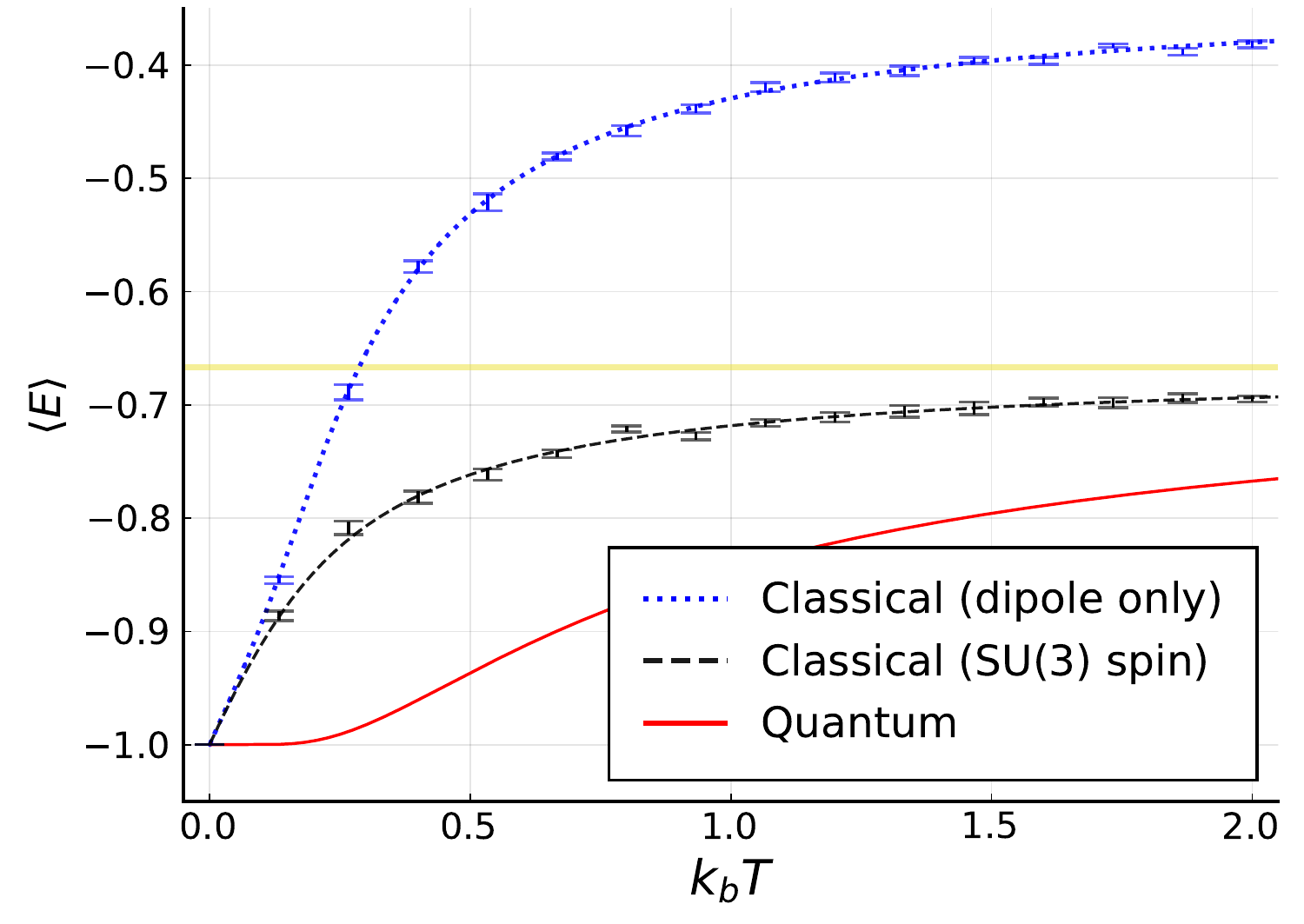}\caption{Energy curves for a single spin with $S=1$ and an easy-axis anisotropy
$D=-1$. The exact quantum mechanical result (solid red) is compared with
the two classical approximations. The first represents spin by a single
dipole moment (blue dots). The second captures both dipole and quadrupole
parts via an SU(3) coherent state (black dashes). At low temperatures, both
classical approximations yield a dipole aligned with the $z$-axis.
Only the approach with SU(3) coherent states converges to the correct
result of $E=2D/3$ (yellow line) at infinite temperature. Statistical
estimates calculated using the stochastic Schrödinger equation (error 
markers) agree with the analytical results.\label{fig:spin1}}

\end{figure}

The three expected energies, $E=-\partial\ln\mathcal{Z}/\partial\beta$,
are
\begin{align}
E_{\mathrm{quantum}} & =\frac{2D}{2+e^{D\beta}}\\
E_{\textrm{SU(3)}} & =\frac{2}{\beta}+\frac{D^{2}\beta}{1-e^{D\beta}+D\beta}\\
E_{\mathrm{dipole}} & =\frac{1}{2\beta}-\frac{\sqrt{D}\,e^{-D\beta}}{\sqrt{\pi\beta}\,\mathrm{erf}(\sqrt{D\beta})}.
\end{align}
Figure~\ref{fig:spin1} shows these energy curves for the case
of $D=-1$. In the zero temperature limit ($\beta\rightarrow\infty$),
both classical approximations yield the correct energy, $E_{\textrm{SU(3)}}=E_{\mathrm{dipole}}=D$.
At high temperatures, however, $E_{\mathrm{dipole}}$ converges to
the incorrect value $D/3$, whereas $E_{\textrm{SU(3)}}$ converges
correctly to $2D/3$. This difference illustrates the importance
of quadrupolar fluctuations present in the SU(3) coherent states, and missing from the dipole-only model.

Error markers in Fig.~\ref{fig:spin1} show statistical estimates for $E_\mathrm{dipole}$ and $E_\textrm{SU(3)}$ obtained by numerical integration of the stochastic Schrödinger dynamics, Eq.~(\ref{eq:stoch_schro}). For this, we used the Heun scheme with normalization, as described in Appendix~\ref{sec:heun}.
At each temperature, energy was estimated as an average over 10 independent numerical trajectories. Each trajectory involved 11k time-steps, each with $\Delta t=0.01$; the first 1k time-steps were discarded, and measurements were taken over the subsequent 10k time-steps. A strong coupling to the thermal bath, $\tilde{\lambda}=1.0$, was selected to approximately optimize decorrelation time.

To estimate $E_\textrm{SU(3)}$ we sampled SU(3) coherent states by integrating Eq.~(\ref{eq:stoch_schro}) with $\mathfrak{H}$ the matrix representation  of the full quantum Hamiltonian defined in Eq.~(\ref{eq:calH_aniso}). Here, each spin operator $\hat{S}^{\alpha}$ was replaced by its representation as a $3\times3$ matrix, corresponding to the fundamental irrep of SU(3). In particular, $\hat{S}^{z}$ was taken to be the diagonal matrix with elements $[1,0,-1]$.

To estimate $E_\mathrm{dipole}$ we sampled normalized spin dipoles from the Boltzmann distribution for the classical Hamiltonian of Eq.~\eqref{eq:H_dipole}. Equation~\eqref{eq:stoch_schro} can again be used, but the construction is a bit subtle. Normalized dipoles map bijectively to SU(2) coherent state vectors via the Bloch sphere. To model dipoles of magnitude $|\mathbf{s}| = 1$, the spin operators $\hat{S}^\alpha$ should be represented by Pauli matrices $\sigma^\alpha$ (note the absence of a 1/2 scaling factor). Time-evolution is then generated by the effective Hamiltonian $\mathfrak{H}_{\mathrm{dipole}}=2Ds^{z} \sigma^{z}$. For more details, see the reference code that accompanies this paper~\cite{RefCode}.

\subsection{Single dimer \label{sec:dimer}}

As a final solvable example, consider a dimer unit of two spin-$1/2$ sites, coupled
by a Heisenberg interaction,
\begin{equation}
\hat{\mathcal{H}}=J\hat{\mathbf{S}}_{1}\cdot\hat{\mathbf{S}}_{2}.
\end{equation}
The quantum Hamiltonian is alternatively written,
\begin{equation}
\hat{\mathcal{H}}=J/4-J|v\rangle\langle v|,\label{eq:calH_decompose}
\end{equation}
where $|v\rangle=(|\uparrow,\downarrow\rangle-|\downarrow,\uparrow\rangle)/\sqrt{2}$
denotes the fully entangled singlet state. The eigenvalues are 
\begin{equation}
\epsilon_{\{1,2,3\}}=J/4,\quad\epsilon_{4}=-3J/4.\label{eq:heis_eig}
\end{equation}
The quantum mechanical partition function is
\[
\mathcal{Z}_{\mathrm{quantum}}=e^{-\beta J/4}(3+e^{+\beta J}),
\]
with associated energy, 
\begin{equation}
E_{\mathrm{quantum}}=\frac{3J(e^{\beta J}-1)}{4(e^{\beta J}+3)}.\label{eq:E_qm_heis}
\end{equation}

The formalism described in this paper allows for two possible classical
approximations. In the traditional approach, one models the spin at
each site with an expected dipole $\mathbf{s}_{j}$ of magnitude $S=1/2$,
yielding the classical Hamiltonian
\begin{equation}
H_{\mathrm{dipoles}}=J\mathbf{s}_{1}\cdot\mathbf{s}_{2}.\label{eq:H_class_heis}
\end{equation}
The corresponding partition function is given by integration over
dipoles $\mathbf{s}_{1}$ and $\mathbf{s}_{2}$ with magnitude $|\mathbf{s}_{j}|=1/2$.
Because the model is invariant to a global rotation, we may replace
$\mathbf{s}_{1}\rightarrow\hat{z}/2$, where $\hat{z}$ is the unit
vector in the $z$-direction. The remaining integral over $\mathbf{s}_{2}$
makes contact with Eq.~(\ref{eq:Z_cl_zeeman_2}) where $B=-J/2$.
The expected energy follows from Eq.~(\ref{eq:E_cl_zeeman}),
\begin{equation}
E_{\mathrm{dipoles}}=\beta^{-1}-\frac{J}{4}\coth(\beta J/4).\label{eq:E_cl_heis}
\end{equation}

An improved classical approximation would treat both spins as a single
``entangled unit'' with $N=4$ levels. That is, a classical configuration
is represented by a four-component complex vector $\mathbf{Z}$, associated
with classical energy,
\begin{equation}
H_{\textrm{SU(4)}}=\mathbf{Z}^{\dagger}\mathfrak{H}\mathbf{Z}.
\end{equation}
The quantum Hamiltonian is exactly represented as a $4\times 4$ matrix, given as a tensor product of Pauli matrices,
\begin{equation}
\mathfrak{H}=\frac{J}{4}\sum_{\alpha=x,y,z}\sigma^{\alpha} \otimes\sigma^{\alpha},\label{eq:frakH_eu}
\end{equation}
The classical
partition function is defined by Eq.~(\ref{eq:Z_gs}) using the eigenvalues
given in Eq.~(\ref{eq:heis_eig}),
\begin{align}
\mathcal{Z}_{\textrm{SU}(4)} & \propto e^{\beta\frac{3J}{4}}\int_{0}^{1}\!\!\mathd y_{1}\int_{0}^{1-y_{1}}\!\!\mathd y_{2}\nonumber \\
 & \quad\times\int_{0}^{1-y_{1}-y_{2}}\!\!\mathd y_{3}\,e^{-\beta J\left(y_{1}+y_{2}+y_{3}\right)}.
\end{align}
The expected energy, $-\partial\ln\mathcal{Z}_{\textrm{SU(4)}}/\partial\beta$,
is
\begin{equation}
E_{\textrm{SU}(4)}=\frac{3}{\beta}-\frac{3}{4}J+\frac{J^{3}\beta^{2}}{2+J\beta(2+J\beta)-2e^{J\beta}}.\label{eq:E_cl_eu}
\end{equation}

\begin{figure}
\includegraphics[width=0.95\columnwidth]{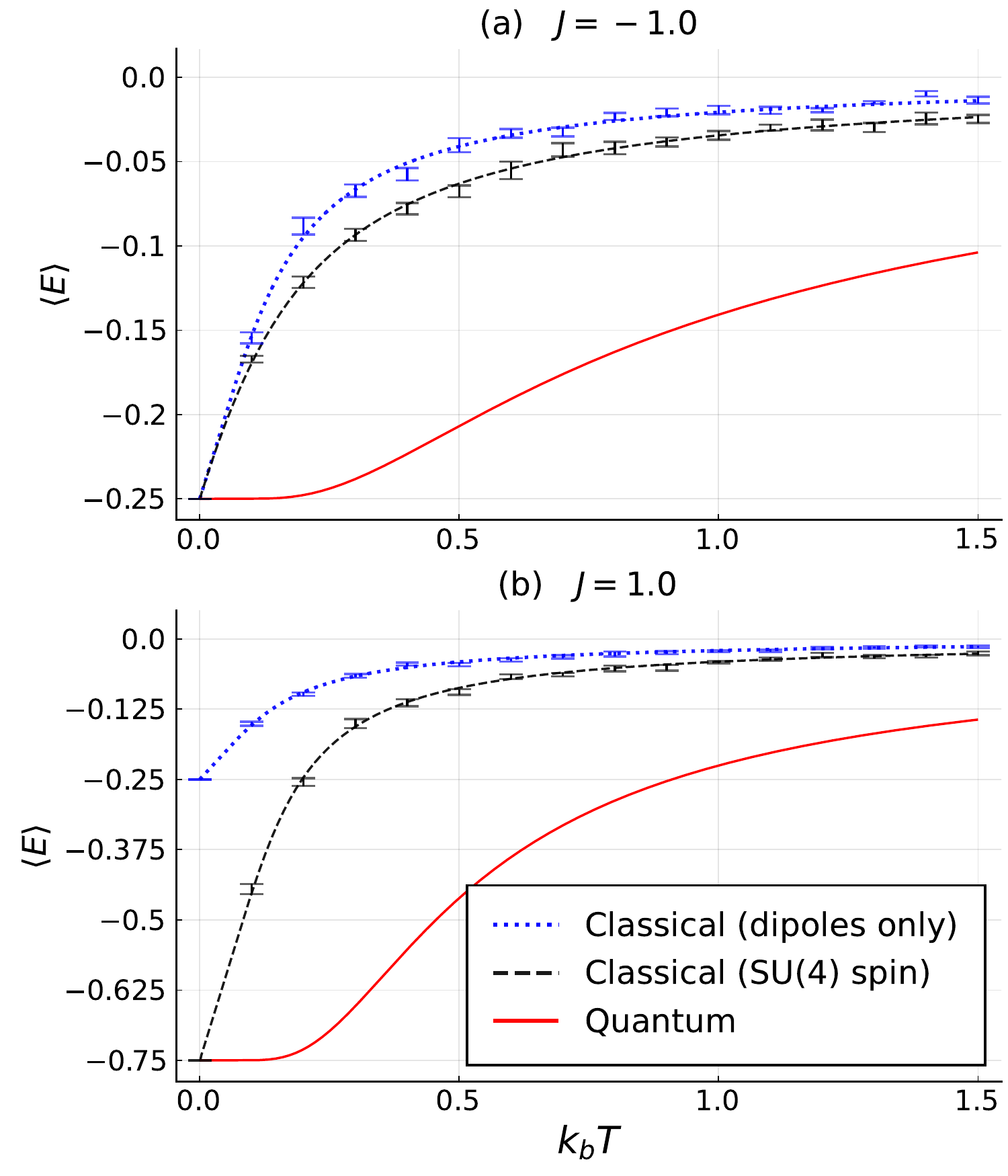}

\caption{Energy curves for two interacting sites with spin $S=1/2$ and (a)
ferromagnetic or (b) antiferromagnetic Heisenberg exchange, $J=\mp1$.
The exact quantum mechanical result (solid red) is compared with the two
classical approximations. The first involves non-entangled dipoles
(blue dots), and the second involves entangled SU(4) coherent states (black dashes).
Only the latter can model the fully entangled singlet state, which is the correct ground state in (b). Statistical estimates calculated
using the stochastic Schrödinger equation (error markers) agree with the analytical results.\label{fig:zeeman}}
\end{figure}

Figure~\ref{fig:zeeman} shows the expected energy as a function
of temperature for ferromagnetic and antiferromagnetic couplings $J$.
In the ferromagnetic case, $J=-1$, both classical approximations
$E_{\mathrm{dipoles}}$ and $E_{\textrm{SU}(4)}$ correctly describe
the ground state, in which the two dipoles are aligned. In the antiferromagnetic
case, $J=+1$, only the treatment involving SU(4) coherent states
correctly captures the true ground state, which is a fully entangled
singlet.

Error markers in Fig.~\ref{fig:zeeman} show statistical estimates obtained by integrating the stochastic Schrödinger dynamics, Eq.~(\ref{eq:stoch_schro}).
In the dipole-only approximation, each site contains a two-component
coherent state, and these evolve according to the local Hamiltonians
$\mathfrak{H}_{1}=Js_{2}^{\alpha}\sigma^{\alpha}/2$ and $\mathfrak{H}_{2}=Js_{1}^{\alpha}\sigma^{\alpha}/2$
where local spin operators $\hat{S}_{j}^{\alpha}$
correspond to the $2\times2$ matrices $\sigma^{\alpha}/2$.
In the treatment using SU(4) coherent states, the stochastic Schrödinger
equation instead uses the $4\times4$ matrix $\mathfrak{H}$ defined
in Eq.~(\ref{eq:frakH_eu}). All other simulation parameters match those used for Fig.~\ref{fig:spin1}.

\subsection{Quenching into CP$^{2}$ skyrmions\label{sec:skyrmion}}

\begin{figure*}
\includegraphics[width=1.0\textwidth]{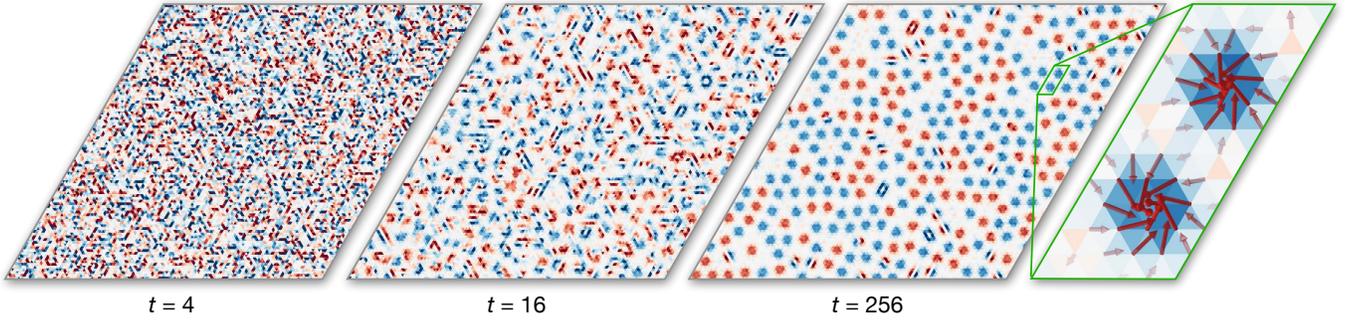}
\caption{
Emergence of CP$^2$ skyrmions in a spin-1 model with competing exchange interactions and an easy-plane anisotropy. The time sequence shows a nonequilibrium quench from infinite to zero temperature using the Langevin dynamics of SU(3) coherent states. The energy damping timescale is $1/\tilde{\lambda} = 10$. All times are measured in units of $\hbar / |J_1|$. The color of each triangular plaquette represents topological charge density for CP$^2$ skyrmions, ranging from positive (red) to negative (blue). The white background corresponds to a quantum paramagnetic phase  with zero expected skyrmion charge, large quadrupole moment, and negligible dipole moment. The far right panel shows a zoom of two well-formed CP$^2$ skyrmions. These skyrmions are long-lived effective particles due to topological protection.\label{fig3}}
\end{figure*}

To illustrate the interesting phenomena that can be studied with the stochastic spin dynamics, Eq.~\eqref{eq:S_GSD} or equivalently Eq.~\eqref{eq:stoch_schro}, we study a quenching process that results in the formation of CP$^2$ skyrmions.  Following Ref.~\onlinecite{zhang2022}, our starting point is a spin-1 model known to contain magnetic field induced skyrmion crystal phases at $T=0$,
\begin{align}
{\hat {\mathcal H}} \! = \! \sum_{i,j} & J_{ij} \left( {\hat S}^x_i {\hat S}^x_j + {\hat S}^y_i {\hat S}^y_j + \Delta {\hat S}^z_i {\hat S}^z_j \right)
\nonumber \\
&-  \sum_{i}  \big[  D  ({\hat S}^z_i )^2 + h {\hat S}^z_i  \big].
\label{hamil}
\end{align}
The exchange interactions $J_{ij}$ are ferromagnetic for nearest-neighbors, with strength $J_1= -1$, and antiferromagnetic for next-nearest neighbors, with strength $J_2= 2 / (1 + \sqrt{5})$. This particular ratio $J_2 / J_1$ favors magnetic spirals with a period of 5 lattice constants. Additionally, the model includes competing exchange and single-ion uniaxial spin anisotropies, $\Delta = 2.6$ and $D = 19$. At the carefully tuned magnetic field of $h=15.35$, the ground state belongs to the SkX-I phase, which describes a crystal of CP$^2$ skyrmions in a quantum paramagnetic background~\cite{zhang2022}. 
As we will demonstrate below, fast temperature quenches can lead instead to a disordered skyrmion configuration. Similar trapping behaviors have been observed experimentally, e.g., in thin films of the non-centrosymmetric material Fe$_{0.5}$Co$_{0.5}$Si~\cite{Yu10}.
In general, CP$^{N-1}$ skyrmions are characterized by an integer topological charge, which may be decomposed as a sum over local contributions $c_\triangle$ defined on triangular plaquettes of nearest neighbor sites; Appendix~\ref{sec:charge} gives the precise definition of this local charge.

Figure~\ref{fig3} shows the time evolution of topological charge $c_\triangle$ for a nonequilibrium temperature quench process. The triangular lattice consists of $100 \times 100$ sites with periodic boundary conditions. Each spin-1 site is associated with a generalized dipolar/quadrupolar spin, or equivalently, an SU(3) coherent state. The initial configuration was totally randomized, corresponding to infinite temperature. Following this, the system was evolved according to the stochastic spin dynamics using the formalism of Eq.~\eqref{eq:stoch_schro}. By selecting a target temperature of $k_\mathrm{B} T = 0$, all Langevin noise was suppressed. Finite coupling with the thermal bath, $\tilde{\lambda} = 0.1$, drained energy from the system at the characteristic time scale $1 / \tilde{\lambda} = 10$ (all times are implicitly measured in units of $\hbar / |J_1|$).
The total time duration of the simulation was $256$, using time steps of $\Delta t = 0.01$ (this scale is set by $1/D$, where the anisotropy strength $D$ is the largest energy scale in the model). The panels in the figure, from left to right, show the progression of CP$^2$ skyrmion formation. At the intermediate time $t=16$ one can observe proto-skyrmion objects, visible as localized regions of positive (red) and negative (blue) charge. The formation of a white background is also apparent, which represents a quantum paramagnetic phase with negligible dipole component. The proto-skyrmions appear oblong, and are typically paired with a partner of either same- or opposite-charge. At the relatively late time $t=256$, we observe well-formed, distinct skyrmions. The effective skyrmion-skyrmion interaction is repulsive, but highly local. Some opposite-charged skyrmion pairs continue to be present at $t=256$, but at a much lower density compared to $t=16$. Also present are ring-shaped objects that contain twice the normal skyrmion charge.  The wall-clock time to simulate this full trajectory ($25,600$ time-steps) is about a minute on a modern laptop computer. See Ancillary file \texttt{T=0\_freezing.mp4} at \href{https://arxiv.org/abs/2209.01265}{[arXiv:2209.01265]} for a movie showing the full quench process.

The true energy-minimizing configuration for this system is believed to be SkX-I, a densely packed CP$^2$ skyrmion crystal~\cite{zhang2022}. At temperatures between roughly $0.06$ and $0.12$, we find the equilibrium phase to be a dilute gas of skyrmions. See the Ancillary file \texttt{T=0.07\_sublimation.mp4} for a movie showing the sublimation of the CP$^2$ skyrmion crystal ground state into this skyrmion gas phase (duration $t=8,500$). Skyrmion creation or annihilation events are present, but relatively rare. Occasional changes in the sign of the skyrmion charge are also present (i.e., switching from red to blue).

Some of this phenomenology has previously been observed for fast quenches of dipole-only spin models, giving rise to metastable CP$^1$ skyrmion configurations~\cite{Lin2016}. Our SU(3) treatment, however, is essential to observe CP$^2$ skyrmions, which can only exist when quadrupolar fluctuations are present.

\section{Conclusions}

We have extended the generalized spin dynamics, Eq.~\eqref{eq:GSD}, with Langevin damping and noise terms. A fluctuation-dissipation theorem ensures that the proposed stochastic spin dynamics samples from the classical Boltzmann distribution. This dynamics can be viewed as a multipolar generalization of the well-known stochastic Landau-Lifshitz equation for spin dipoles. Such a generalization is necessary to model a wide class of spin systems with $S > 1/2$ and strong single-ion anisotropy due to crystal field effects. For example, the simulation methods described here underpin recent theoretical modeling of the $S=1$ antiferromagnet Ba$_2$FeSi$_2$O$_7$~\cite{Do22}. The framework of SU($N$) states can also be useful for modeling entangled units of strongly coupled sites. For example, in Sec.~\ref{sec:dimer} we considered a spin-1/2 dimer with antiferromagnetic dimer, and demonstrated how a classical model involving SU(4) coherent states correctly describes the fully entangled singlet ground state.

To facilitate numerical simulations, we developed a mathematical mapping from the stochastic spin dynamics, Eq.~\eqref{eq:S_GSD}, to an equivalent formulation in terms of SU($N$) coherent states, Eq.~\eqref{eq:stoch_schro}. As a demonstration of this framework, we studied a nonequilibrium process for a spin-1 model that gives rise to a long-lived, metastable liquid of CP$^2$ skyrmions.
The generalized Langevin spin dynamics presented in this work could also be used to simulate of the motion of CP$^{N-1}$ skyrmions driven by spin polarized currents. 

The code used to reproduce the numerical results in this work is available online~\cite{RefCode}. A general purpose framework for developing and simulating SU($N$) spin models is provided by the \texttt{Sunny} open-source package~\cite{Sunny}.

\begin{acknowledgments}
The authors thank Shi-Zeng Lin and Leandro Chinellato for helpful discussions. D.D. and C.D.B.~acknowledge support from U.S. Department of Energy, Office of Science, Office of Basic Energy Sciences, under award DE-SC-0018660. 
This work was performed in part at Aspen Center for Physics, which is supported by National Science Foundation grant PHY-1607611. K.B.~acknowledges support from the Center of Materials Theory as a part of the Computational Materials Science (CMS) program, funded by the U.S. Department of Energy, Office of Basic Energy Sciences.
\end{acknowledgments}

\appendix

\section{Fluctuation-dissipation theorem for the stochastic spin dynamics\label{sec:fluctuation_dissipation}}

Here we derive a fluctuation-dissipation theorem for stochastic spin
dynamics, Eq.~(\ref{eq:S_GSD}), following previous arguments~\cite{Hasegawa80,Skubic08}.

A Lie-Poisson system has the general form

\begin{equation}
\frac{dy}{dt}=\mathcal{B}\nabla H.\label{eq:liepoisson2}
\end{equation}
where $H(y)$ is the Hamiltonian and $\mathcal{B}$ has the matrix
elements

\begin{equation}
\mathcal{B}_{ij}=C_{ij}^{k}y_{k}.\label{eq:B_def}
\end{equation}
In this appendix, summation over repeated Roman indices is implied.
The symbol $C_{ij}^{k}$ denotes the structure constants for an arbitrary
Lie algebra $\mathfrak{g}$. From the antisymmetry of the Lie bracket,
$C_{ij}^{k}=-C_{ji}^{k}$, or equivalently, $\mathcal{B}^{T}=-\mathcal{B}$.
For semi-simple Lie algebras, one can select generators such that
$C_{ij}^{k}$ is also antisymmetric in its third index $k$~\cite{Modin20}.
We will assume this total antisymmetry,
\begin{equation}
C_{ij}^{k}=-C_{ji}^{k}=-C_{kj}^{i}.\label{eq:C_antisym}
\end{equation}

The generalized spin dynamics stated in Eq.~(\ref{eq:GSD}) has the
form of a Lie-Poisson system. If the system had only a single site
$j=1$, then $y=\mathbf{n}_{1}$ would be the dynamical vector, $C_{\alpha\beta}^{\gamma}=f_{\alpha\beta\gamma}$
would be the structure constants for SU($N$), and $\mathcal{B}_{\alpha\beta}=f_{\alpha\beta\gamma}n_{1}^{\gamma}$
would be the matrix that implements the generalized cross product,
i.e., $\mathcal{B}\nabla_{1}H=-\mathbf{n}_{1}\star\nabla_{1}H$.
For a system with multiple sites, the vector $y=[\mathbf{n}_{1},\dots\mathbf{n}_{L}]$
grows to include all spins and the matrix $\mathcal{B}$ becomes block
diagonal, $\mathcal{B}_{(j\alpha),(k\beta)}=\delta_{jk}f_{\alpha\beta\gamma}n_{j}^{\gamma}$.
The diagonal blocks of $\mathcal{B}$ still implement the generalized
cross product, now for each site independently. This matrix can also
be written $\mathcal{B}_{(j\alpha),(k\beta)}=C_{(j\alpha),(k\beta)}^{(\ell\gamma)}y_{(\ell\gamma)}$,
involving the structure constants $C_{(j\alpha),(k\beta)}^{(\ell\gamma)}=\delta_{jk}\delta_{j\ell}f_{\alpha\beta\gamma}$,
which are nonzero unless all three sites are equal, $j=k=\ell$. It follows
that $C_{(j\alpha),(k\beta)}^{(\ell\gamma)}$ inherits the total antisymmetry
of $f_{\alpha\beta\gamma}$. In other words, Eq.~(\ref{eq:C_antisym})
holds where the letters $(i,j,k\dots)$ are understood to
denote both site \emph{and} generator indices.

The stochastic spin dynamics of Eq.~(\ref{eq:S_GSD}) takes the form,

\begin{equation}
\frac{\mathd y}{\mathd t}=\mathcal{B}(\xi+\nabla H+\lambda\mathcal{B}\nabla H).\label{eq:langevin}
\end{equation}
The noise term $\xi$ appears multiplicatively, and should be integrated
using the Stratonovich calculus. Each component $\xi_{i}(t)$ is Gaussian
white noise, with first and second moments
\begin{align}
\left\langle \xi_{i}(t)\right\rangle  & =0\\
\left\langle \xi_{i}(t)\xi_{j}(t')\right\rangle  & =2D\delta_{ij}\delta(t-t').
\end{align}

By defining
\begin{align}
\mathcal{A} & =\mathcal{B}\left(\nabla H+\lambda\mathcal{B}\nabla H\right),\label{eq:A_def}
\end{align}
we can split the Langevin equation into its deterministic and stochastic
parts,
\begin{equation}
\frac{\mathd y}{\mathd t}=\mathcal{A}+\mathcal{B}\xi,\label{eq:langevin_general}
\end{equation}
The corresponding Fokker-Planck equation describes the time-evolution
of the probability distribution $P(y)$ under the Langevin dynamics~\cite{Brown63},
\begin{align}
\frac{\partial P}{\partial t} & =-\frac{\partial}{\partial y_{i}}\left(\mathcal{A}_{i}P\right)+D\frac{\partial}{\partial y_{i}}\left[\mathcal{B}_{ik}\frac{\partial}{\partial y_{j}}\left(\mathcal{B}_{jk}P\right)\right].\label{eq:fokker1}
\end{align}
Chapter 9 of Ref.~\cite{vanKampen07} reviews the mapping from a
Langevin to a Fokker-Planck equation.

Our aim is to show that the Boltzmann distribution
\begin{equation}
P(y)=\frac{1}{Z}e^{-\beta H(y)}.\label{eq:boltzmann_y}
\end{equation}
is a stationary point of Fokker-Planck equation, provided that we
select the noise magnitude 
\begin{equation}
D=\lambda k_\mathrm{B}T=\lambda/\beta.\label{eq:D_def}
\end{equation}

Using Eqs.~(\ref{eq:B_def}) and~(\ref{eq:C_antisym}), we find
\begin{equation}
\frac{\partial}{\partial y_{j}}\mathcal{B}_{jk}=C_{jk}^{j}=0.\label{eq:dB_dy_0}
\end{equation}
Using the antisymmetry $\mathcal{B}_{jk}=-\mathcal{B}_{kj}$,
\begin{equation}
\frac{\partial}{\partial y_{j}}\left(\mathcal{B}_{jk}P\right)=-\mathcal{B}_{kj}\frac{\partial P}{\partial y_{j}}.
\end{equation}
Substitution into~(\ref{eq:fokker1}) yields,
\begin{align}
\frac{\partial P}{\partial t} & =-\frac{\partial}{\partial y_{i}}\left(\mathcal{A}_{i}P\right)-D\frac{\partial}{\partial y_{i}}\left(\mathcal{B}_{ik}\mathcal{B}_{kj}\frac{\partial P}{\partial y_{j}}\right).\label{eq:fokker2}
\end{align}

Focusing on the first term, we substitute from Eq.~(\ref{eq:A_def})
to find
\begin{align}
\frac{\partial}{\partial y_{i}}\left(\mathcal{A}_{i}P\right) & =\frac{\partial}{\partial y_{i}}\left(\mathcal{B}_{ij}\frac{\partial H}{\partial y_{j}}P+\lambda\mathcal{B}_{ik}\mathcal{B}_{kj}\frac{\partial H}{\partial y_{j}}P\right).\label{eq:dAP_dy}
\end{align}

The derivative of the Boltzmann distribution in Eq.~(\ref{eq:boltzmann_y})
is
\begin{equation}
\frac{\partial P}{\partial y_{j}}=\frac{\partial}{\partial y_{j}}\frac{e^{-\beta H}}{Z}=-\beta\frac{\partial H}{\partial y_{j}}P.
\end{equation}
With this result, Eq.~(\ref{eq:dAP_dy}) becomes
\begin{equation}
\frac{\partial}{\partial y_{i}}\left(\mathcal{A}_{i}P\right)=-\beta^{-1}\frac{\partial}{\partial y_{i}}\left(\mathcal{B}_{ij}\frac{\partial P}{\partial y_{j}}+\lambda\mathcal{B}_{ik}\mathcal{B}_{kj}\frac{\partial P}{\partial y_{j}}\right).
\end{equation}
Substitution into~(\ref{eq:fokker2}) yields
\begin{equation}
\frac{\partial P}{\partial t}=\beta^{-1}\frac{\partial}{\partial y_{i}}\mathcal{B}_{ij}\frac{\partial P}{\partial y_{j}}+\left(\lambda\beta^{-1}-D\right)\frac{\partial}{\partial y_{i}}\left(\mathcal{B}_{ik}\mathcal{B}_{kj}\frac{\partial P}{\partial y_{j}}\right).
\end{equation}
The second term vanishes when the noise magnitude $D$ is selected
as in Eq.~(\ref{eq:D_def}). Differentiating the remaining term yields
\begin{equation}
\frac{\partial P}{\partial t}=\beta^{-1}\left(\mathcal{B}_{ij}\frac{\partial^{2}P}{\partial y_{i}\partial y_{j}}+\frac{\partial\mathcal{B}_{ij}}{\partial y_{i}}\frac{\partial P}{\partial y_{j}}\right).
\end{equation}
The first term vanishes because $\mathcal{B}_{ij}$ is antisymmetric
whereas $\partial^{2}P/\partial y_{i}\partial y_{j}$ is symmetric.
The second term vanishes by Eq.~(\ref{eq:dB_dy_0}). We conclude
that the Boltzmann distribution is stationary, $\partial P/\partial t=0$,
completing our demonstration of the fluctuation-dissipation theorem.

\section{Gaussian distributed random Hermitian matrices\label{sec:noise}}

Here we explore the properties of random Hermitian matrices with Gaussian
distributed elements.

This paper has focused on the Lie group SU($N$) in the fundamental
representation, with generators $T^{\alpha}$ satisfying $\mathrm{tr}\,T^{\alpha}T^{\beta}=\tau\delta_{\alpha\beta}$
for some magnitude $\tau$. These generators span the space of $N\times N$
\emph{traceless} Hermitian matrices. It is useful to consider the
extension to the Lie group U($N$), which has Hermitian generators
$E^{\alpha}$ that are not all traceless. We will continue to impose
an orthonormality condition,

\begin{equation}
\mathrm{tr}\,E^{\alpha}E^{\beta}=\tau\delta_{\alpha\beta}.
\end{equation}
One possibility is to reuse $E^{\alpha}=T^{\alpha}$ for $\alpha=1,\dots,N^{2}-1$,
and include an additional generator $E^{\alpha=N^{2}}=\sqrt{\tau}I$.

A random, Gaussian distributed Hermitian matrix is

\begin{equation}
\mathcal{X}=x^{\alpha}E^{\alpha},\label{eq:calX}
\end{equation}
where $x^{\alpha}$ for $\alpha=1,\dots,N^{2}$ are Gaussian random
variables satisfying
\begin{equation}
\langle x^{\alpha}\rangle=0,\quad\langle x^{\alpha}x^{\beta}\rangle=\delta_{\alpha\beta}.
\end{equation}

\subsection{Basis independence of the distribution}

A key property of $\mathcal{X}$ is that its distribution is independent
of the choice of generators $E^{\alpha}$. To demonstrate this, consider
some other generators $E'^{\alpha}$, also orthonormal, $\mathrm{tr}\,E'^{\alpha}E'^{\beta}=\tau\delta_{\alpha\beta}.$

We can always find real coefficients $R_{\alpha\beta}$ that transform
between the two bases,
\begin{equation}
E'^{\alpha}=R_{\alpha\beta}E^{\beta}.\label{eq:T_transform-1}
\end{equation}
Using the orthonormality conditions, it follows
\begin{align}
\tau\delta_{\alpha,\beta} & =\mathrm{tr}E'^{\alpha}E'^{\beta}\nonumber \\
 & =R_{\alpha\gamma}R_{\beta\delta}\mathrm{tr}E^{\gamma}E^{\delta}\nonumber \\
 & =\tau(RR^{T})_{\alpha,\beta}.
\end{align}
In other words, $R$ is an orthogonal matrix, $RR^{T}=I$.

Consider a new random matrix,
\begin{equation}
\mathcal{X}'=x^{\alpha}E'^{\alpha},
\end{equation}
where the coefficients $x_{\alpha}$ are shared with Eq.~(\ref{eq:calX}).
Substitution of Eq.~(\ref{eq:T_transform-1}) yields
\begin{equation}
\mathcal{X}'=(x^{\alpha}R_{\alpha\beta})E=x'^{\alpha}E^{\alpha}.
\end{equation}
The coefficients $\mathbf{x}$ and $\mathbf{x}'=R\mathbf{x}$ share
the same Gaussian distribution because $R$ is norm preserving.
It follows that the random matrices $\mathcal{X}$ and $\mathcal{X}'$ also share the same distribution.

Given any unitary $U$, we can create a new set of orthonormal generators,
$E'^{\alpha}=UE^{\alpha}U^{\dagger}$. The distribution of $\mathcal{X}$
is invariant under the unitary transformation $\mathcal{X}\rightarrow\mathcal{X}'=U\mathcal{X}U^{\dagger}$.

\subsection{Sampling random matrices}

We have seen that the distribution of $\mathcal{X}$ is independent
of the choice of $N^2$ orthonormal generators $E^{\alpha}$. A valid selection is the set:
\begin{enumerate}
\item $\sqrt{\frac{\tau}{2}}\left(e_{ab}+e_{ba}\right)$ for $a>b$
\item $\im\sqrt{\frac{\tau}{2}}\left(e_{ab}-e_{ba}\right)$ for $a>b$
\item $\sqrt{\tau}e_{aa}$ for $a=1,\dots,N$,
\end{enumerate}
where $e_{ab}$ denotes the matrix with a 1 in the $(a,b)$th entry,
and 0 elsewhere. Substituting these $E^{\alpha}$ into Eq.~(\ref{eq:calX})
yields random matrix elements
\begin{equation}
\mathcal{X}=\sqrt{\tau}\,\left[\begin{array}{cccc}
r_{11} & h_{21}^{\ast} & \cdots & h_{N1}^{\ast}\\
h_{21} & r_{22} &  & \vdots\\
\vdots &  & \ddots & h_{N,N-1}^{\ast}\\
h_{N1} & \cdots & h_{N,N-1} & r_{NN}
\end{array}\right].\label{eq:calX_dist}
\end{equation}
Each $h_{ab}$ for $a>b$ is an independent complex Gaussian random
variable with zero mean and unit variance. Similarly, $r_{aa}$ is
an independent real Gaussian random variable with zero mean and unit
variance.

An alternative but equivalent construction is $\mathcal{X}=\sqrt{\tau/2}\,(A+A^{\dagger})$,
where every matrix element $A_{ab}$ is an independently sampled complex
Gaussian random variable with zero mean and unit variance.

\subsection{Sampling random matrix-vector products}

Let $\mathbf{v}$ denote any normalized vector, with $|\mathbf{v}|=1$.
We will characterize the distribution of the random vector $\mathcal{X}\mathbf{v}$.

It will be convenient to work with the special vector
\begin{equation}
\mathbf{e}=\left[\begin{array}{c}
1\\
0\\
\vdots\\
0
\end{array}\right].
\end{equation}
Select some unitary transformation $U$ that satisfies $U^{\dagger}\mathbf{v}=\mathbf{e}$.
Such a transformation exists because $|\mathbf{v}|=|\mathbf{e}|$.
Since $\mathcal{X}$ is statistically invariant under a unitary transformation,
the distribution of the vector $\mathcal{X}\mathbf{v}$ is identical
to that of $U\mathcal{X}U^{\dagger}\mathbf{v}$. Consider
the latter form. Taking $\mathcal{X}$ to be sampled as in Eq.~(\ref{eq:calX_dist}),
we find that
\begin{equation}
\mathcal{X}U^{\dagger}\mathbf{v}=\sqrt{\tau}\,\left[\begin{array}{c}
r\\
h_{2}\\
\vdots\\
h_{N}
\end{array}\right],
\end{equation}
where each $h_{a}$ is a complex Gaussian variable with unit variance,
and $r$ is a real Gaussian variable with unit variance. Introducing
an additional complex Gaussian component $h_{1}$ (independent of
$r$), this may be written,
\begin{equation}
\mathcal{X}U^{\dagger}\mathbf{v}=\sqrt{\tau}\,[(I-\mathbf{e}\mathbf{e}^{\dagger})\mathbf{h}+r\mathbf{e}].
\end{equation}
Noting that $U\mathbf{e}=\mathbf{v}$ and $U(I-\mathbf{e}\mathbf{e}^{\dagger})=(I-\mathbf{v}\mathbf{v}^{\dagger})U$,
we find
\begin{equation}
U\mathcal{X}U^{\dagger}\mathbf{v}=\sqrt{\tau}\,\left[(I-\mathbf{v}\mathbf{v}^{\dagger})U\mathbf{h}+r\mathbf{v}\right].
\end{equation}
The random vector $\mathbf{h}'=U\mathbf{h}$ shares the same Gaussian distribution
as $\mathbf{h}$ because unitary transformations are norm preserving. Therefore we may effectively
replace $U\mathbf{h}\rightarrow\mathbf{h}$.

Our final result is that the random vector $\mathcal{X}\mathbf{v}$
can be sampled using
\begin{equation}
\mathcal{X}\mathbf{v}=\sqrt{\tau}\,\left[(I-\mathbf{v}\mathbf{v}^{\dagger})\mathbf{h}+r\mathbf{v}\right],\label{eq:gaussian_times_vec}
\end{equation}
where $\mathbf{h}$ is a complex random Gaussian vector with second
moments $\langle h_{a}^{\ast}h_{b}\rangle=\delta_{ab}$, and $r$
is a real random Gaussian with unit variance.

\subsection{Gaussian white noise appearing in the stochastic spin dynamics}

Recall that Eq.~(\ref{eq:calX}) constructs $\mathcal{X}$ as a random
linear combination of the orthonormal generators $E^{\alpha}$ of
U($N$). The stochastic spin dynamics involves instead the traceless
generators $T^{\alpha}$ of SU($N$), to which we may associate a
new random matrix distribution,
\begin{equation}
\tilde{\mathcal{X}}=x^{\alpha}T^{\alpha},
\end{equation}
where the sum now runs from $\alpha=1$ to $N^{2}-1$. Given a sample
for $\mathcal{X}$, we may construct a sample for $\tilde{\mathcal{X}}$
by removing the trace,
\begin{equation}
\tilde{\mathcal{X}}=\mathcal{X}-(\mathrm{tr}\mathcal{X}/N)I.
\end{equation}
To check this, recall that the distribution for $\mathcal{X}$ is
independent of the choice of generators $E^{\alpha}$; the selection
$\{E^{1},\dots,E^{N^{2}}\}=\{T^{1},\dots,T^{N^{2}-1},\sqrt{\tau}\,I\}$
establishes the identity.

Using $\mathbf{v}^{\dagger}\mathbf{v}=1$, an immediate corollary
of Eq.~(\ref{eq:gaussian_times_vec}) is
\begin{equation}
(I-\mathbf{v}\mathbf{v}^{\dagger})\mathcal{X}\mathbf{v}=\sqrt{\tau}\,(I-\mathbf{v}\mathbf{v}^{\dagger})\mathbf{h},
\end{equation}
The left-hand side is invariant under the substitution $\mathcal{X}\rightarrow\mathcal{X}+cI$,
for arbitrary $c$. This allows us to effectively replace $\mathcal{X}\rightarrow\tilde{\mathcal{X}}$,
yielding the identity
\begin{equation}
(I-\mathbf{v}\mathbf{v}^{\dagger})\tilde{\mathcal{X}}\mathbf{v}=\sqrt{\tau}\,(I-\mathbf{v}\mathbf{v}^{\dagger})\mathbf{h}.\label{eq:gaussian_claim1}
\end{equation}

This result can be restated in the language of Gaussian white noise
processes, which will make contact with the stochastic spin dynamics,
Eq.~(\ref{eq:S_GSD}). Here we will focus on a single site and drop
the site index $j$. The analog of $x^{\alpha}$ are the noise components
$\xi^{\alpha}$, which are characterized by first and second moments
{[}cf. Eqs.~(\ref{eq:xi1}) and~(\ref{eq:xi2}){]},
\begin{align}
\left\langle \xi^{\alpha}(t)\right\rangle  & =0\\
\left\langle \xi^{\alpha}(t)\xi^{\beta}(t')\right\rangle  & =2D\delta_{\alpha\beta}\delta(t-t').
\end{align}
We also introduce a complex\emph{ }Gaussian white noise,
\begin{align}
\langle\zeta_{a}(t)\rangle & =0\\
\left\langle \zeta_{a}^{\ast}(t)\zeta_{b}(t')\right\rangle  & =2\tau D\delta_{ab}\delta(t-t').
\end{align}
Gaussian white noise, integrated over an arbitrary interval, yields
an ordinary Gaussian random variable. For example,
\begin{equation}
\int_{0}^{\Delta t}\zeta_{a}(t)\mathd t=\sqrt{\tau}ch_{a},
\end{equation}
where $c=\sqrt{2D\Delta t}$, and $h_{a}$ are complex Gaussian variables
defined by the moments $\langle h_{a}\rangle=0$ and $\langle h_{a}^{\ast}h_{b}\rangle=\delta_{ab}$.

The analog of $\tilde{\mathcal{X}}$ is the matrix noise {[}cf. Eq.~(\ref{eq:frakX}){]},
\begin{equation}
\mathfrak{X}(t)=\xi^{\alpha}(t)T^{\alpha}.
\end{equation}
Integration yields,
\begin{equation}
\int_{0}^{\Delta t}\mathfrak{X}(t)\mathd t=T^{\alpha}\int_{0}^{\Delta t}\xi^{\alpha}(t)\mathd t=c\tilde{\mathcal{X}}.
\end{equation}
Collecting these results, we observe that
\begin{equation}
(I-\mathbf{v}\mathbf{v}^{\dagger})\mathfrak{X}\mathbf{v}=\sqrt{\tau}\,(I-\mathbf{v}\mathbf{v}^{\dagger})\boldsymbol{\zeta},
\end{equation}
reduces exactly to Eq.~(\ref{eq:gaussian_claim1}) when integrated
over any interval. Therefore the equation is correct in general, for
any normalized $\mathbf{v}$.

In our application to Langevin spin dynamics, we are working with
a stochastic differential equation for $\mathbf{Z}(t)$ that contains
multiplicative noise, Eq.~(\ref{eq:s-gsd-intermediate}). We wish
to apply the identity,
\begin{equation}
(I-\mathbf{Z}\mathbf{Z}^{\dagger})\mathfrak{X}\mathbf{Z}=\sqrt{\tau}\,(I-\mathbf{Z}\mathbf{Z}^{\dagger})\boldsymbol{\zeta},\label{eq:gaussian_noise_final}
\end{equation}
where $\mathbf{v}\rightarrow\mathbf{Z}(t)$ now evolves stochastically
in time. Again, we can justify Eq.~(\ref{eq:gaussian_noise_final})
by integrating both sides over an arbitrarily small interval. The
validity of this procedure depends crucially on the Stratonovich interpretation
of Eq.~(\ref{eq:s-gsd-intermediate})~\cite{vanKampen07}.

\section{Numerical integration\label{sec:heun}}

To numerically integrate the stochastic Schrödinger dynamics, Eq.~(\ref{eq:stoch_schro}),
a good scheme is second-order Heun followed by a normalization
step~\cite{Mentink10}. To express this procedure, it is helpful to write the stochastic dynamics in a compact form,
\begin{equation}
\frac{\mathd}{\mathd t}\mathbf{Z}_{j}=\mathcal{A}_{j}[\mathbf{Z}]+\mathcal{B}_{j}[\mathbf{Z}]\boldsymbol{\mathbf{\zeta}}_{j},
\end{equation}
involving the drift term $\mathcal{A}_{j}=-\im\,P_{j}(1-\im\,\tilde{\lambda})\,\mathfrak{H}_{j}\mathbf{Z}_{j}$, and
the noise scaling factor $\mathcal{B}_{j}=-\im\,P_{j}$,
where $P_{j}=I-\mathbf{Z}_{j}\mathbf{Z}_{j}^{\dagger}$ was defined in Eq.~\eqref{eq:P_def}.  The bracket
notation implies functional dependence on all sites in the system,
e.g., $\mathcal{A}_{j}[\mathbf{Z}]=\mathcal{A}_{j}(\mathbf{Z}_{1},\dots,\mathbf{Z}_{L})$.

One forward Euler integration time-step gives a predictor for the
update,
\begin{equation}
\mathbf{Z}_{j}^{(1)}=\mathbf{Z}_{j}+\Delta t\mathcal{A}_{j}[\mathbf{Z}]+\sqrt{\Delta t}\,\mathcal{B}_{j}[\mathbf{Z}]\mathbf{g}_{j}.
\end{equation}
Loosely speaking, the random complex vector $\sqrt{\Delta t}\,\mathbf{g}_{j}$
represents the integral of $\boldsymbol{\zeta}_{j}$ over the integration
time-step $\Delta t$. Its components are Gaussian distributed with
zero mean and second moment
\begin{equation}
\langle g_{j,a}^{\ast}g_{k,b}\rangle=2\tilde{\lambda}k_\mathrm{B}T\delta_{jk}\delta_{ab}.
\end{equation}
In practice, to calculate each component \textbf{$g_{j,a}$,} we sample
complex Gaussian random variables $h_{j,a}$ with zero mean and unit
variance and then rescale, $g_{j,a}=\sqrt{2\tilde{\lambda}k_\mathrm{B}T}\,h_{j,a}.$
Note that the real and imaginary parts of $h_{j,a}$ are individually
Gaussian distributed, with variance $1/2$.

Given this predictor $\mathbf{Z}^{(1)}$, the Heun method uses a corrector
step,
\begin{align*}
\mathbf{Z}_{j}^{(2)} & =\mathbf{Z}_{j}+\Delta t\left(\frac{\mathcal{A}_{j}[\mathbf{Z}]+\mathcal{A}_{j}[\mathbf{Z}^{(1)}]}{2}\right)\\
 & \quad\quad+\sqrt{\Delta t}\,\left(\frac{\mathcal{B}_{j}[\mathbf{Z}]+\mathcal{B}_{j}[\mathbf{Z}^{(1)}]}{2}\right)\mathbf{g}_{j}.
\end{align*}
Finally we employ the normalization
\begin{equation}
\mathbf{Z}'_{j}=\mathbf{Z}_{j}^{(2)}/|\mathbf{Z}_{j}^{(2)}|,
\end{equation}
to ensure an exact unitary evolution. The final update rule over the
time-step $\Delta t$ is $\mathbf{Z}_{j}\rightarrow\mathbf{Z}_{j}'$.

For small $\Delta t$, the Heun scheme converges correctly to the
solution of Eq.~(\ref{eq:stoch_schro}) under the Stratonovich interpretation,
for which the noise term is evaluated at the midpoint\emph{ }of the
time-step~\cite{Mentink10}. Note that forward Euler, without a
corrector step, would converge to an incorrect limit (this would be
the Itô interpretation).

\section{Skyrmion charge\label{sec:charge}}

Here we review the general definition of CP$^{N-1}$ skyrmions, which are localized topological defects. The topological charge of a skyrmion is  formally defined for a continuous field of SU($N$) coherent states $\mathbf Z(\mathbf r) \in \mathrm{CP}^{N-1}$, where $\mathbf r$ denotes position in the 2D plane. Each coherent state $\mathbf{Z}$ may be interpreted as a normalized, $N$-component complex vector, but any two coherent states that differ only by a complex phase are identified as the same element in $\mathrm{CP}^{N-1}$. Assuming that the field of coherent states is uniform at infinity, $|\mathbf{r}| \rightarrow \infty$, the spatial plane may be compactified onto the 2-sphere, $S^2$. That is, we may view $\mathbf{Z}(\mathbf r)$ as a continuous map from $S^2$ to CP$^{N-1}$. The associated homotopy group $\pi_2(\mathrm{CP}^{N-1})\cong \mathbb{Z}$ characterizes the topologically distinct integer winding numbers. In physics, this winding number is known as the (baby) skyrmion charge.

From $\mathbf{Z}(\mathbf r)$ follows the generalized spin components $n^{\alpha}(\mathbf{r})$, defined as in Eq.~(\ref{eq:n_def}), and the color field
\begin{equation}
    \mathfrak{n}(\mathbf{r}) = n^{\alpha}(\mathbf{r})T^\alpha.
    \label{color_field}
\end{equation}
Up to a proportionality constant, the CP$^{N-1}$ skyrmion charge is defined as
\begin{equation}
C \propto -\im \int \textrm{tr}\,\left(\mathfrak{n}\left[\partial_x\mathfrak{n} , \partial_y \mathfrak{n}\right]\right) \, \mathd \mathbf{r},\label{eq:charge}
\end{equation}
where $\partial_x$ and $\partial_y$ denote partial derivatives with respect to the Cartesian components of the position vector $\mathbf{r}$. The constant of proportionality is $N$-dependent, and should be selected so that the possible values for $C$ are the set of integers, $\mathbb Z$. 
The traditional  CP$^1$ skyrmions  appearing in condensed matter physics are composed of dipoles, or equivalently, SU(2) coherent states. The spin-1 system defined in Eq.~\eqref{hamil} involves instead SU(3) coherent states.

To discretize the skyrmion charge $C$ onto the lattice, consider a triangular plaquette $\triangle = \langle jkl \rangle$ comprised of three nearest-neighbor sites. Using a generalization of Stokes theorem, the area integral of charge density over the plaquette becomes a line integral along the triangle boundary, $j \rightarrow k \rightarrow l \rightarrow j$, oriented clockwise. One can interpolate the color field between any two nearest neighbor sites using the CP$^{N-1}$ geodesic. The final result for the CP$^{N-1}$ skyrmion charge on the plaquette is,
\begin{equation}
    c_\triangle = \frac{1}{2\pi} \left( \gamma_{jl} + \gamma_{lk} + \gamma_{kj}\right).\label{eq:charge_density}
\end{equation}
where $\gamma_{kj}=\textrm{arg}(\mathbf{Z}_k^\dagger \mathbf{Z}_j)$ is the Berry connection on the bond $j\rightarrow k$, and $N$ may be arbitrary. The total skyrmion charge on a lattice is given by the sum over oriented plaquettes, $C = \sum_\triangle c_\triangle$. For a finite lattice with periodic boundary conditions, this sum over plaquettes is exactly integer.

\bibliographystyle{apsrev4-2}
\bibliography{refs}

\end{document}